\documentclass[prb,twocolumn,preprintnumbers,amsmath,amssymb]{revtex4}

\usepackage{epsfig,amsfonts}

\def\be{\begin{equation}}
\def\ee{\end{equation}}

\def\bea{\begin{eqnarray}}
\def\eea{\end{eqnarray}}

\def\e{\epsilon}

\def\a{\alpha}
\def\b{\beta}
\def\g{\gamma}
\def\s{\sigma}

\def\ben{\begin{enumerate}}
\def\een{\end{enumerate}}

\begin{document}

\title{Dynamical correlation functions of the mesoscopic pairing model}
\author{Alexandre Faribault${}^1$, Pasquale Calabrese${}^{2}$ and  
Jean-S\'ebastien Caux${}^{3}$}
\affiliation{$^1$Physics Department, ASC and CeNS,
Ludwig-Maximilians-Universit\"at, 80333 M\"unchen, Germany}
\affiliation{$^{2}$Dipartimento di Fisica dell'Universit\`a di Pisa and  
INFN,
56127 Pisa, Italy}
\affiliation{$^3$Institute for Theoretical Physics, Universiteit van  
Amsterdam,
1018 XE Amsterdam, The Netherlands}

\date{\today}

\begin{abstract}
We study the dynamical correlation functions of the Richardson pairing  
model (also known as the reduced or discrete-state BCS model) in the
canonical ensemble.  
We use the Algebraic Bethe Ansatz  formalism, which gives exact expressions 
for the form factors of the most important observables. 
By summing these form factors over a relevant set of states, we obtain 
very precise estimates of the correlation functions, as confirmed by 
global sum-rules (saturation above 99\% in all cases considered).  
Unlike the case of many other Bethe Ansatz solvable theories, 
simple two-particle states are sufficient to achieve such saturations,
even in the thermodynamic limit.
We provide explicit results at half-filling, and discuss their finite-size scaling behavior.
\end{abstract}

\maketitle

\section{Introduction}
Possibly the most remarkable incipient property associated to a fermionic
gas is its instability to the pairing phenomenon.  Under an arbitrarily
weak attractive force, which can originate in such a simple process as
coupling to phonons, the gas will develop an instability towards the formation
of Cooper pairs, this process forming the basis of the BCS theory
of superconductivity \cite{bcs-57}, with further remarkable consequences 
such as the Meissner and Josephson effects.  Within BCS theory, single-particle
excitations are suppressed by the superconducting gap, which is obtained from the solution of a variational
ansatz for the wavefunction defined in the grand-canonical ensemble 
(electron number is by construction not conserved anymore)
and in the thermodynamic limit (since a continuous energy band is assumed).  

While these mean-field approaches successfully describe the experimental
features of the traditional bulk superconductors, recent experiments have also
considered metallic nanograins \cite{exp,dr-01} in which the level spacing is
finite and of the same order as the superconducting gap, and in which some Coulomb
blockade effects could occur in view of the finite charging energy of the grain.  
These studies open the door to many interesting further questions not answerable
within BCS theory, and require
mesoscopic effects to be encompassed back into the model.  One way to approach
this problem is to use the so-called
reduced BCS model, defined by the Hamiltonian 
\be
H_{BCS}=\sum^N_{\stackrel {\a=1}{\sigma=+, -}}
\frac{\e_\a}2  c^\dagger_{\a\sigma}c_{\a\sigma}
  -g\sum^N_{\a,\b=1} c^\dagger_{\a+}c^\dagger_{\a-}
  c_{\b-}c_{\b+},\label{BCSH}
\ee
which was introduced by Richardson in the early 1960's in the context  
of nuclear physics  \cite{rs-62}.
The model describes (pseudo) spin-$1/2$ fermions
(electrons, nucleons, etc\dots) in a shell of doubly degenerate single
particle energy levels with energies $\e_\a/2$, $\a = 1, \dots N$.
$c_{\a,\s}$ are the fermionic
annihilation operators, $\s = +,-$ labels the degenerate time
reversed states (i.e. spin or isospin) and
$g$ denotes the effective pairing coupling constant.
Despite its simplified character (the interaction couples all levels uniformly),
the model does have a number of advantages as compared to BCS theory.
First of all, it can be solved within the canonical ensemble (fixed number of
electrons), a situation which is relevant for isolated nanograins.  
Second, and rather remarkably for an exactly solvable model, 
it remains solvable for an arbitrary choice of parameters, and can
thus provide quantitative predictions for various situations obtained by considering 
various choices of the set of energy levels $\e_\a$ (both their number, and their 
individual value), coupling $g$, and filling.  Besides mesoscopic superconductivity,
this model and its solution also find applications in other fields (see 
the reviews [\onlinecite{dps-04,dh-03}] for some applications outside of condensed matter physics). 

The nature of the electronic states in a metallic nanograin can conceivably be
probed in a number of different experiments.  Electronic transport through such a grain
could be studied by attaching either metallic or superconducting leads.  The observable
I-V characteristics or Josephson currents would be theoretically obtainable from 
correlation functions within the grain.  Such correlations, however, are not easily obtainable
from the basic exact solution of the model, which focuses on wavefunctions but does not allow
to make direct contact with the dynamics of observables.  The history of the study of correlations in the
Richardson model is however already rather rich.  Richardson himself in 1965 \cite{r-65} derived a  first
exact expression for static correlation functions.
In a significant development, Amico and Osterloh \cite{ao-02} proposed  
a new method  to write  down such correlations explicitly but with results 
limited to system sizes of up to 16 particles.
A major simplification was then proposed by Zhou et al. \cite{zlmg-02,lzmg-03}:  
using the Algebraic Bethe Ansatz (ABA) and  the Slavnov
formula for scalar products of states \cite{s-89}, they managed to obtain
the static correlation functions as sums over $N_p^2$ determinants of $N_p\times N_p$ matrices. 
These expressions have been further simplified by us in a previous publication, where they were
evaluated numerically for a particular choice of the energy levels $\e_\a$  \cite{fcc-08}
allowing to describe the crossover from mesoscopic to  macroscopic physics, going
beyond previous results limited to fewer particles \cite{mff-98,ao-02}.
We mention that a different approach, valid in the case of highly degenerate $\e_\a$,
is also available \cite{sbb-08}.

Despite all these developments, up to now none of these approaches has been adapted and used to
calculate dynamical correlation functions, important to quantify the response of a physical 
system to any realistic experimental probe.  
In this paper we use a method that mixes integrability and numerics and is similar to that
used by some of us to study the dynamical correlation functions \cite{sc} and 
entanglement entropy \cite{afc-09} in spin-chains, as well as dynamical correlations in Bose 
gases \cite{1dbg}.
In short, the ABACUS method \cite{c-09} consists in using the exact knowledge of
the form factors of physical observables as determinants of matrices whose entries are the 
unknown Richardson rapidities. For any given state, these rapidities are calculated 
by solving the Richardson equations. The corresponding form factors are then evaluated. 
Finally we need to sum over all these contributions, but this can be done by searching for the states 
in the Hilbert space that contribute more significantly to the correlation function. 
This is done by optimizing properly the scanning of the Hilbert space \cite{c-09} and the accuracy of the 
result is kept under control by checking the values of the sum-rules. 
However, we will see that for the Richardson model, this scanning is particularly easy since 
the two-particle states dominate the sum even when increasing the number of particles (in 
strong contrast with what happens for other models \cite{sc,1dbg}). This property is clearly connected with the mean-field character of the model in the thermodynamic limit  and the subsequent suppression of quantum fluctuations. However at finite (sufficiently low) number of particles the effect of quantum fluctuations can be revealed by small, but maybe measurable, multi-particle channels. 
We mention that another variation on the ABACUS logic has recently been used for the calculation of out-of-equilibrium 
observables in the pairing model subjected to a quantum quench, 
but in that case the class of excitations contributing was much wider \cite{fcc-09,fcc-09p}.

The paper is organized as follows.
In Sec. \ref{secmod} we discuss the model and its general properties.
In Sec. \ref{secaba} we recall the Algebraic Bethe Ansatz approach to the model 
and we proceed to some simplifications of the determinant expressions.
In Sec. \ref{srsec} we introduce the sum rules for the considered correlation functions and we
present the selection rules for the form factors that will help in the numerical computation by reducing the 
number of intermediate states we have to sum over. 
In Sec. \ref{seccorr} all the correlation functions are explicitly calculated at half-filling.
We report our main conclusions and discuss open problems for future investigation
in Sec. \ref{concl}. 
Appendix \ref{secgs} reports the technical details on how to solve the Richardson equations  
for any excited state while Appendix \ref{scaling} looks at the details of the strong coupling expansion of a studied correlation function in order to explain its scaling behavior.

\section{The Model}
\label{secmod}
As it is written in Eq. (\ref{BCSH}), singly- as well as doubly-occupied levels are allowed.  
The interaction however couples only doubly-occupied levels among themselves, and due to the 
so-called blocking effect \cite{rs-62,dr-01}, unpaired particles
completely decouple from the dynamics and behave as if they were free.
We will denote the total number
of fermions as $N_f$, and the total number of pairs as $N_p$.
Due to level blocking, we will thus only consider $N_f=2 N_p$ paired  
particles in $N$ unblocked levels, keeping in mind that we could reintroduce blocked levels later 
if needed in the actual phenomenology desired.  In terms of pair annihilation and creation operators
\be
b_\a= c_{\a-}c_{\a+}\, \qquad b_\a^\dagger=  
c^\dagger_{\a+}c^\dagger_{\a-}\,,
\ee
the Hamiltonian is
\be
H= \sum^N_{\a=1}
\e_\a  b^\dagger_\a b_\a -g\sum^N_{\a,\b=1} b^\dagger_\a b_\b\,,
\ee
and $n_\a= 2b_\a^\dagger b_\a$ is the number of
particles in level $\a$.

The pair creation and annihilation operators satisfy the commutation  
relations
\be
[b_\a,b_\b^\dagger]= \delta_{\a\b}(1-2b^\dagger_\a b_\a)\,,\qquad
[b_\a,b_\b]=[b_\a^\dagger,b_\b^\dagger]=0\,.
\ee
The term $2b^\dagger_\a b_\a$ in the first commutator makes the model  
different from free
bosons and therefore non-trivial.

Using the pseudo-spin realization of electron pairs
$S^z_\a= b^{\dagger}_\a b_\a -1/2$, $S^-_\a= b_\a$,  
$S^+_\a=b^\dagger_\a$,
the BCS Hamiltonian becomes (up to a constant)
\be
H=\sum^N_{\a=1}
\e_\a S^z_\a -g \sum^N_{\a,\b=1} S^+_\a S^-_\b \,.
\label{spinH}
\ee
The operators $S_\a^{\pm,z}$ obey the standard $su(2)$ spin algebra and so
the Hamiltonian (\ref{spinH}) describes a spin-$1/2$ magnet with  
long-range
interaction for the $XY$ components in a site-dependent longitudinal 
magnetic field
$\e_\a$.
Such a magnetic Hamiltonian is known in the literature as a
Gaudin magnet \cite{g-book}.
An important relation is
\be
S^\pm_\a S^{\mp}_\a=S_\a^2-(S_\a^z)^2\pm S_\a^z\,.
\label{comm}
\ee

\subsection{Grand-canonical BCS wavefunction}

In the grand-canonical (GC) ensemble the ground state wavefunction is  
the BCS
variational ansatz
\be
|GS\rangle=\prod_\a (u_\a +e^{i\phi_\a} v_\a b^\dagger_\a)|0\rangle\,,
\qquad u_\a^2+v_\a^2=1\,,
\ee
where the variational parameters $u_\a$ and $v_\a$ are real and  
$\phi_\a$ is
a phase which, it turns out, must be $\a$-independent.
$|GS\rangle$ is not an eigenstate of the particle number operator $N_f$
and the average condition $\langle N_f \rangle= \bar{N}_f$ determines
the GC chemical potential. Likewise, the commonly used definition
\be
\Delta_{GC}= 2 g \sum_\a \langle b_\a \rangle=
2 g \sum_\a u_\a v_\a e^{i\phi_\a}\,,
\label{DelGC}
\ee
for the superconducting gap makes sense only in a GC ensemble, since
$\langle b_\a \rangle$ is zero when evaluated at fixed particle number.
The variational parameters are obtained as
\be
v_\a^2=\frac12\left[1-\frac{\e_\a-\mu}{\sqrt{(\e_\a- 
\mu)^2+|\Delta_{GC}|^2}}
\right]\,,
\label{vjBCS}
\ee
where $\mu$ is the GC chemical potential.

\subsection{Canonical description and Richardson solution}

The exact solution (i.e. the full set of eigenstates and eigenvalues) of the Hamiltonian (\ref{BCSH}) 
in the canonical ensemble was derived by Richardson \cite{rs-62}.
The model can be encompassed into the framework on integrable models \cite{crs-97} and is
tractable by means of algebraic
methods \cite{aff-01,dp-02,zlmg-02,lzmg-03}.
We review here only the main points of this solution.

In the ABA, eigenstates are contructed by applying raising operators on
a so-called reference state (pseudovacuum).  We here choose the  
pseudovacuum
(in the pseudo-spin representation) to be fully polarized along the $-\hat{z}$  
axis
  \be
S_\a^z |0\rangle = - \frac{1}{2} |0\rangle\,, \quad \forall \ \a.
\label{pv}
\ee
In the pair representation, this state thus corresponds to the Fock vacuum.
Eigenstates with $N_p$ pairs are then characterized by $N_p$  
spectral parameters
(rapidities) $w_j$, and take the form of Bethe wavefunctions
\be
|\{w_j\}\rangle=\prod_{k=1}^{N_p} \mathcal{C}(w_k) |0\rangle\,.
\label{RICHWF}
\ee
The operators $\mathcal{C}$, together with operators $\mathcal{A},  
\mathcal{B}, \mathcal{D}$
defined as
\bea
&&\!\!\mathcal{A}(w_k) = \frac{-1}{g} + \sum_{\alpha=1}^N  
\frac{S^z_\a}{w_k - \e_\a}, \hspace{0.2cm}
\mathcal{B}(w_k)=\sum_{\a=1}^N \frac{S_\a^-}{w_k - \e_\a}, \nonumber \\
&&\!\!\mathcal{C}(w_k)=\sum_{\a=1}^N \frac{S_\a^+}{w_k - \e_\a},  
\hspace{0.2cm}
\mathcal{D}(w_k) = \frac{1}{g} - \sum_{\alpha=1}^N \frac{S^z_\a}{w_k -  
\e_\a}
\label{inverse_problem}
\eea
obey the Gaudin algebra, which is the quasi-classical limit of the  
quadratic Yang-Baxter
algebra associated to the $gl(2)$ invariant $R$-matrix (we refer the  
readers to [\onlinecite{lzmg-03}]
for details).

The wavefunctions (\ref{RICHWF}) are eigenstates of the transfer matrix,
and thus of the Hamiltonian (\ref{BCSH}), when the parameters $w_j$  
satisfy
the Richardson equations
\be
-\frac1g=\sum_{\a=1}^N\frac1{w_j - \e_\a}-\sum_{k\neq j}^{N_p}  
\frac2{w_j-w_k}\,\quad
j=1,\dots, N_p\,.
\label{RICHEQ}
\ee
Throughout the paper we will refer with latin indices to the rapidities
and with greek ones to the energy levels.
The total energy of a Bethe state is, up to a constant,
\be
E_{\{ w\} }=  \sum_j w_j \,.
\label{enB}
\ee  
For a
given $N$ and $N_p$ the number of solutions of the Richardson equations is
$\binom{N}{N_p}$, and coincides with the
dimension of the Hilbert space of $N_p$ pairs distributed into  
$N$
different levels, i.e. the solutions to the Richardson equations give
all the eigenstates of the model.

Note that one could also use the pseudovacuum to be the state fully polarized along the $\hat{z}$ axis (instead of$-\hat{z}$) This leads to slight differences in the expressions, which one can resolve by comparing references [\onlinecite{zlmg-02}] and [\onlinecite{lzmg-03}].

\section{Correlation functions and Algebraic Bethe Ansatz}
\label{secaba}

We are interested in the dynamical correlation functions of the form
\be
G_{O_{\a\b}}(t)=\frac{\langle GS| O_\a^\dag(t) O_\b(0)|GS \rangle}{
\langle GS| GS \rangle}\,,
\ee
at fixed number of pairs $N_p$. Here $O_\a$ stands for a `local' operator in the Heisenberg 
picture (i.e. depending on a single energy level $\a$). 
We consider $O_\a$ equal to $S^z_\a$ or $S^\pm_\a$.

By inserting the complete set of states $|\{ w \}\rangle$, with $\{ w \}$ a set of $M_w$ rapidities solution 
to the Richardson equations, and using the time evolution of the eigenstate,
we can rewrite the dynamical correlation function as the sum
\begin{multline}
G_{O_{\a\b}}(t)=\\
\sum_{\{ w\}} \frac{\langle \{ w\} |O_\a|GS\rangle^* \langle \{ w\} |O_\b|GS\rangle 
e^{i \omega_w t }}{
\langle GS| GS \rangle\langle \{ w\}| \{ w\} \rangle}\,,
\end{multline}
where  form factors and norms are obtainable by Algebraic Bethe Ansatz \cite{zlmg-02,lzmg-03} and
will be discussed in the next subsection.
The frequency $\omega_w$ is just $\omega_w= E_w-E_{GS}-\mu(M_w-N_{p})$, 
where the energies $E_\mu$ and $E_{GS}$ are given in equation (\ref{enB}). 
$\mu$ is the chemical potential (needed only for  $S^\pm$) and $M_{w}$ is the 
number of rapidities of the state $|\{ w\}\rangle$.
More easily, we can write $\omega_w=E_w-E_{0,w}$, where $E_{0,w}$ is the lowest energy state
with the same number of rapidities as $|\{ w \} \rangle$.
Notice that for $S_z$ correlation function, we only need states with $M_w=N_{p}$, and so the 
chemical potential term is absent while for $\langle S^+S^-\rangle$, we need only states with $N_{p}-1$
rapidities.

The most relevant physical observables are clearly global ones, when the sum over the 
internal energy levels is performed. 
We will consider diagonal and global correlation functions, whose static counterparts for $S^\pm$ 
are the diagonal \cite{dzgt-96} and off-diagonal order parameters. 
We consider in the following the three global correlators
\bea
G_{zz}^d(t)&=&\sum_{\a=1}^N 
\frac{\langle GS| S^z_\a(t) S^z_\a(0)|GS \rangle}{\langle GS| GS \rangle}\,,\\
G_{+-}^d(t)&=&\sum_{\a=1}^N 
\frac{\langle GS| S^+_\a(t) S^-_\a(0)|GS \rangle}{\langle GS| GS \rangle}\,,\\
G_{+-}^{od}(t)&=&\sum_{\a,\b=1}^N 
\frac{\langle GS| S^+_\a(t) S^-_\b(0)|GS \rangle}{\langle GS| GS \rangle}\,.
\eea
We will mainly study them in frequency space, since their structure is less complicated, and we 
will only plot a few examples in real time. 

We stress that while the level-resolved correlators depend strongly on the choice of the 
energy levels $\e_\a$, for the global ones it is expected that most of the qualitative features 
and several quantitative ones are not affected by the choice of the model.
Thus our results, even if obtained for a specific choice of $\e_\a$, should display the main 
features of the dynamical correlation functions for a wide variety of Richardson models.

\subsection{Algebraic Bethe Ansatz and Form Factors}

The starting point to calculate correlation functions with the  
Algebraic Bethe
Ansatz is having a representation for the scalar products
of two generic states defined by $N_p$ rapidities ($N_p$ Cooper pairs)
\be
\langle \{w\}|\{v\}\rangle=
\langle 0| \prod^{N_p}_{b=1} \mathcal{B}(w_b) \prod^{N_p}_{a=1}  
\mathcal{C}(v_a) |0\rangle\,,
\ee
when at least one set of parameters (e.g. $w_b$ but not $v_a$)
is a solution to the Richardson equations.  Following standard  
notations, $\mathcal{C}$ is the conjugate of the operator  
$\mathcal{B}$.
Such a representation exists, and is known as the Slavnov formula  
\cite{s-89},
which for the case at hand specifically reads \cite{zlmg-02}
\bea
\langle \{w\}|\{v\}\rangle &=&
\frac {\prod^{N_p}_{a \neq b}(v_b-w_a)}
{\prod _{b <a} (w_b -w_a) \prod _{a <b} (v_b -v_a)}
\nonumber\\ && \times {\rm det}_{N_p} J(\{ v_a \}, \{ w_b \})\,,
\eea
where the matrix elements of $J$ are given by
\bea
J_{ab} &=& \frac {v_b -w_b}{v_a -w_b}
\left ( \sum^N_{\a=1} \frac {1}
{(v_a -\e _\a)(w_b -\e_\a)}\right. \nonumber\\ && \left.-2\sum _{c\neq  
a}^{N_p} \frac {1}{(v_a-v_c)(w_b -v_c)}
\right ),
\eea
from which the norms of states simply follow from $v\to w$ as
$ ||\{v\}||^2=\det_{N_p} G$ with a Gaudin matrix
\be
G_{ab}=
\begin{cases}\displaystyle
\sum_{\b=1}^N \frac1{(v_a-\e_\b)^2}-2\sum_{c\neq  
a}^{N_p}\frac1{(v_a-v_c)^2}\quad &
a=b\,,\\ \displaystyle
\frac2{(v_a-v_b)^2}& a\neq b\,,
\end{cases}
\label{Gaudin}
\ee
recovering Richardson's old result \cite{r-65}.

The key point is that any form factor of a local spin operator between  
two Bethe eigenstates can
be represented via (\ref{inverse_problem}) as a scalar product with
one set, e.g.  $\{v\}$ not satisfying the Bethe equations, for which
Slavnov's formula is applicable.
This has been explicitly worked out in Ref. [\onlinecite{zlmg-02}].
For $\{w\}$,$ \{v\}$ containing respectively $N_p+1$ and $N_p$  
elements, the
non-zero form factors are:
\bea
\langle \{w\}|S^-_\a|\{v\}\rangle =
  \langle \{v\}|S^+_\a|\{w\}\rangle =
\nonumber\\ \frac {\prod^{N_p+1}_{b=1} (w_b - \e_\a)} {\prod  
^{N_p}_{a=1} (v_a - \e_\a)}
  \frac { {\rm det}_{N_p +1} T (\a, \{w\}, \{ v\})}
{\prod _{b > a} (w_b -w_a) \prod _{b <a} (v_b -v_a)}\,,
\label{matels-}
\eea
and, for both $\{w\}$ and $\{v\}$ containing $N_p$ rapidities
\bea
\langle \{w\}|S^z_\a|\{v\}\rangle =
\prod^{N_p}_{a=1} \frac {(w_a - \e_\a)} {(v_a - \e_\a)} \nonumber\\  
\times
\frac { {\rm det}_{N_p} \left (\frac12 T_z(\{w\}, \{v\})
- Q (\a, \{ w\}, \{ v\}) \right )}
{\prod _{b > a} (w_b -w_a) \prod _{b <a} (v_b -v_a)}\,,
\eea
with the matrix elements of $T$ given by ($b < N_p+1$)
\bea
T_{ab}(\a) =&&
\prod ^{N_p+1}_{{c \neq a}} (w_c - v_b)
\left ( \sum^N_{\a=1} \frac1{(v_b -\e _\a)(w_a -\e_\a)}
\right.\nonumber\\ && \left.-2\sum _{c \neq a} \frac1{(v_b-w_c)(w_a  
-w_c)} \right )\,, \nonumber\\
T_{a N_p+1}(\a)  && =  \frac {1}{(w_a -\e_\a)^2}, \ \
Q_{ab}(\a) = \frac {\prod _{c \neq b} (v_c-v_b)} {(w_a-\e_\a)^2}.
\nonumber
\eea
Above, $T_z$ is the $N_p \times N_p$ matrix
obtained from $T$ by deleting the last row and column and replacing  
$N_p+1$ by
$N_p$ in the matrix elements. Here it is assumed that both $\{ v_a \}$  
and
$\{ w_b \}$ are solutions to Richardson's Bethe equations.
However, the results are still valid for $S^\pm_\a$ if only
$\{ w_b \}$ satisfy the Bethe equations.

When approaching a bifurcation point (see App. \ref{secgs}) in the solutions of the Richardson equations, 
some individual terms in the sum defining the matrix elements of $T$ 
tend to diverge (because $w_b\to \e_\a$ for 
some $b$ and $\a$). Those divergences cancel out when the sum is taken, but they can still lead to 
large numerical inaccuracies. By using the Richardson equations, it is possible to eliminate such potentially problematic terms and rewrite the matrix $T$  as
\be
T_{ab} =
\frac{2\displaystyle\prod ^{N_p}_{{c \neq a}} (w_c - v_b)}{w_a-v_b}
\left[\sum _{c \neq b} \frac1{(v_b-v_c)}-\sum _{c \neq a} \frac1{(v_b-w_c)} 
\right], 
\ee
where potentially diverging terms have been removed.
We stress once again that such a formula holds only if {\it both} $\{w_a\}$ and $\{v_a\}$ 
are solutions to the Richardson equations with the same $g$.
This expression has been obtained before in Ref. \onlinecite{fcc-09}.

\section{Sum rules and selection rules}
\label{srsec}

The standard way to assess the accuracy of a numerical calculation in which we discard part of the 
states consists in using sum rules, e.g. summing of all the contributions independently of the energy of the 
intermediate state. When summing over all states, we always get static quantities, and, in the present 
case, they can be obtained by very simple considerations. 

For $G_{zz}^d$ we have the sum rule
\bea
&&\sum_{\a=1}^N\sum_{\{ v\}} 
\frac{ |\langle \{ v\} |S^z_\a|GS\rangle|^2}{
\langle GS| GS \rangle\langle \{ v\} | \{ v\} \rangle}= \nonumber
\sum_{\a=1}^N
\frac{ \langle GS |(S^z_\a)^2|GS\rangle}{
\langle GS| GS \rangle}\\&&=
\langle (S^z)^2 \rangle= \frac{N}4\,.
\label{srzz}
\eea
Only the $\binom{N}{N_p}$ states with the same number of rapidities as the ground state
contribute to this correlation function.
Instead for $G_{+-}^d$ we have
\be
\sum_{\a=1}^N
\frac{ \langle GS |S^+_\a S^-_\a|GS\rangle}{
\langle GS| GS \rangle}=\frac{N}2+ \langle S^z \rangle =\frac{N}2\,,
\label{sr+-}
\ee
as easily shown by using Eq. (\ref{comm}).
Here, only the $\binom{N}{N_p-1}$ states with one less 
rapidity than the ground state contribute to this correlation and to the similar one containing off-diagonal terms $G_{+-}^{od}$. In this last case we have
\be
 \sum_{\a,\b=1}^N \frac{\langle GS |S^+_\a S^-_\b|GS\rangle}{\langle GS| GS \rangle}
 \equiv \Psi_{od}\,,
 \label{sr+-od}
\ee
that is the off-diagonal order parameter, which can be obtained by the solution of the Richardson equations
for the ground state and using the Hellmann-Feynman theorem \cite{fcc-08}. 

We will see in the following that the two-particle states will give most of the contribution to the correlation 
functions, always saturating the sum-rules to more than $99\%$ accuracy. We show in the following 
subsections that some selection rules imply that only two-particle states have non-zero contribution to the 
correlation functions for $g=0$ and for $g\to\infty$.
Although at intermediate couplings this set of states does not give 100\% saturation of the sum rules, 
these two limits clearly give insight as to why they remain extremely dominant in every regime.

\subsection{Weak coupling regime}

In the non-interacting $g = 0$ limit, the fixed $N_p$ eigenstates are quite naturally described by placing 
the $N_p$ Cooper pairs (flipped pseudo-spins) in any of the $\binom{N}{N_p}$ possible sets of $N_p$  energy levels picked from the $N$ available ones. 
This translates into a representation in terms of rapidities given by setting the $N_p$ rapidities to be 
strictly equal to the energies $\epsilon_\alpha$ of the $N_p$ levels occupied by a pair. Since 
\be
\lim_{u\to \epsilon_\alpha} {\cal C}(u) = \lim_{u\to \epsilon_\alpha} \frac{S^+_i}{u-\epsilon_\alpha} \,,
\ee 
the states built in such way will have diverging norms and form factors, but it remains possible to describe 
the limit correctly because in the ratio of form factors and norm the two divergences cancel. Since any of these states is an eigenvector of every $S^z_\alpha$ operators, at $g=0$ the only contributions to the $S^z$ correlations come from the ground state to ground state form factor $\left<GS\right|S^z_\alpha\left|GS\right>$. In a perturbative expansion \cite{sild-01} in $g$, it is easy to see that at first order, the corrections to the ground state comes only from states  $\left|\left\{ w\right\}=\left\{ \epsilon_{\alpha'_1} ... \epsilon_{\alpha'_{N_p}}\right\}\right>$  differing from it by at most one rapidity (in the $g\to 0$ limit). They constitute the full set of two-particle states, obtained by creating a "hole" and a "particle", i.e. moving a single Cooper pair (rapidity) in the ground state to any of the available unoccupied states.

At $g=0$, the $\left<\left\{ v\right\}\right|S^-_\alpha\left|GS\right>$ form factors are non-zero whenever $\left|\left\{ v\right\}=\left\{ \epsilon_{\alpha_1} ... \epsilon_{\alpha_{N_p-1}}\right\}\right>$ is obtained by removing a single rapidity from the $N_p$ pairs ground state $\left|GS=\left\{ w_1=\epsilon_{1},  ... \ w_{N_p}= \epsilon_{N_p}\right\}\right>$. These states can also all be thought of as two-particle states in the $N_p-1$ pairs sector, since they can all be generated by moving a single rapidity in the $N_p-1$ ground state.

In the specific case of half filling ($N_p=N/{2}$), treating every two-particle excitation means that only ${N^2}/{4}$ states are needed, out of the full $\binom{N}{N/2}$ dimensional Hilbert space. Quite naturally, when a non-zero coupling is included these states might not be sufficient anymore, but as will be shown in the next two sections, for $g\to \infty$, only this set of states is once again needed to compute every non-zero form factor of local spin operators. 

\subsection{Strong coupling ($g \to \infty$) regime}
\label{strongg}

Yuzbashyan et al. \cite{yba-03,yba-05} showed that the solutions to the Richardson equations are such that 
in the $g \to \infty$ limit, a number $N_r$ of the rapidities will diverge as $w_i \approx C_i g + \mathcal{O}(g^0)$.
The coefficients $C_i$ are given by the $N_r$ roots of the appropriate Laguerre polynomial  \cite{yba-03}
\bea
L_{N_r}^{-1-N-2(N_r-N_p)}(C_i) =0. 
\label{laguerre}
\eea
In the infinitely large coupling limit, the impact of the diverging rapidities is well defined. 
In fact, $\displaystyle \lim_{u\to\infty} {\cal C}(u) \propto \sum_\alpha S^+_\alpha = S^+_{\mathrm{tot}}$ 
is the total spin raising operator. 
In this limit, the coupling term $g\displaystyle\sum_{\alpha\beta} S^+_\alpha S^-_\beta = g S^+_{tot} S^-_{tot}$
completely dominates the Hamiltonian and so its eigenstates become eigenstates of the $\mathbf{S}_{tot}^2$ 
operator too. 
Through simple energetic consideration \cite{yba-03,sg-06}, one can then easily show that the number of 
diverging rapidities is related to the eigenvalue $J (J+1)$ of $\mathbf{S}_{tot}^2$ by the relation
\bea
N_r = J + N_p -\frac{N}{2}.
\label{corr}
\eea
A state defined by $N_p$ rapidities, of which $N_r$ are infinite, therefore belongs to a subspace 
of the total Fock space defined by eigenvalues of $S^z_{tot}$ and $\mathbf{S}_{tot}^2$ given 
by quantum numbers $m = -N/2 + N_p$ and $J = N_r-N_p + {N}/{2}$. 
This allows us to derive explicit strong-coupling selection rules for the various form factors used in this work. 

Since a given eigenstate, in this limit, is built out of a linear superposition of the various spin states with fixed $J$ and $m$, it can be decomposed onto the joint eigenbasis of spin $\alpha$ combined with the various multiplets emerging from the addition of the remaining $N-1$ spins $\frac{1}{2}$. 

For a given (fixed degeneracy index $k$) $N-1$ spins multiplet with magnitude $J'$, one can write the highest weight state as
\be
\left|k,J=J'+\frac{1}{2},m=J'+\frac{1}{2} \right>=\left|\uparrow_\alpha\right> \otimes \left|k , J',m'=J' \right>\,.
\ee
The $n$-times repeated action of $S^-_{tot} = S^-_\alpha+S^-_{N-1}$ on this state will generate the eigenstates 
\bea
\left|k,J'+\frac{1}{2},m'+\frac{1}{2} - n \right>&=&C^n_1\left|\uparrow_\alpha\right> \otimes \left|k , J',J'-n \right> \nonumber\\&
+& C^n_2 \left|\downarrow_\alpha\right> \otimes \left|k , J',J'-n+1 \right>,\nonumber
\eea
where we do not need to explicitly specify the Clebsh-Gordan coefficients. 
Combining this multiplet with a spin $\frac{1}{2}$ also gives rise to a second set of states given by 
$J= J' -{1}/{2}$ and the corresponding allowed values of $m \in \left\{-J, -J+1, ... J-1, J\right\}$. 
These states are easily constructed by making them orthogonal to the previously found ones, i.e. 
\begin{multline}
\left|k,J'-\frac{1}{2},m'+\frac{1}{2} - n \right>=-C^n_2\left|\uparrow_\alpha\right> \otimes \left|k , J',J'-n \right> 
\\+ C^n_1 \left|\downarrow_\alpha\right> \otimes \left|k , J',J'-n+1 \right>. 
\end{multline}

Any general state with fixed $J$ and $m$ can therefore have contributions coming from $J+{1}/{2}$ 
or $J-{1}/{2}$ multiplets of the $N-1$ excluded spins, i.e.:
\bea
\sum_{k} A_{k}\left|k,J,m \right> &=& \sum_k \left(B^k_1 \left|\uparrow_\alpha\right> \otimes \left|k , J-\frac{1}{2},m-\frac{1}{2} \right>  \right. \nonumber\\ && +B^k_2 \left|\downarrow_\alpha\right> \otimes \left|k , J-\frac{1}{2},m+\frac{1}{2} \right>
\nonumber\\ &+&
B^k_3 \left|\uparrow_\alpha\right> \otimes \left|k , J+\frac{1}{2},m-\frac{1}{2} \right>  \nonumber\\ &+& \left.B^k_4 \left|\downarrow_\alpha\right> \otimes \left|k , J+\frac{1}{2},m+\frac{1}{2} \right>\right).
\eea
Given this  form, it is trivial to see that the application of $S^z_\alpha$ or $S^-_\alpha$ on any of these states will 
result in  
\bea
S^z_\alpha\sum_{k} A_{k}\left|k,J,m \right> &\propto& \sum_k \left[B^k_1 \left|\uparrow_\alpha\right> \otimes \left|k , J-\frac{1}{2},m-\frac{1}{2} \right>  \right. \nonumber\\ && -B^k_2 \left|\downarrow_\alpha\right> \otimes \left|k , J-\frac{1}{2},m+\frac{1}{2} \right>
\nonumber\\ && +
B^k_3 \left|\uparrow_\alpha\right> \otimes \left|k , J+\frac{1}{2},m-\frac{1}{2} \right>  \nonumber\\ && \left.-B^k_4 \left|\downarrow_\alpha\right> \otimes \left|k , J+\frac{1}{2},m+\frac{1}{2} \right>\right]\,,
\nonumber\\
S^-_\alpha\sum_{k} A_{k}\left|k,J,m \right> &\propto& \sum_k \left[B^k_1 \left|\downarrow_\alpha\right> \otimes \left|k , J-\frac{1}{2},m-\frac{1}{2} \right>  \right.\nonumber\\ &&  \left. +
B^k_3 \left|\downarrow_\alpha\right> \otimes \left|k , J+\frac{1}{2},m-\frac{1}{2} \right> \right].\nonumber
\eea
Consequently, the form factors for $S^z_\alpha$ 
\bea
\left(\sum_{k'} A'_{k'}\left<k',J'',m'' \right|\right)S^z_\alpha\left(\sum_{k} A_{k}\left|k,J,m \right>\right),
\eea
can exclusively be non-zero if $J'' \in \left\{J-1,J,J+1\right\}$ and $m''=m$.

Similarly, the $S^-_\alpha$ form factor
\bea
\left(\sum_{k'} A'_{k'}\left<k',J'',m'' \right|\right)S^-_\alpha\left(\sum_{k} A_{k}\left|k,J,m \right>\right),
\eea
is non-zero only if  $J''  \in \left\{J-1,J,J+1\right\}$ and $m'' = m -1$.

 Since, as we pointed out earlier, the value of $J$ is related to the number of diverging rapidities, 
these selection rules translate into selection rules for the total number of rapidities and the number of diverging rapidities. 

The $S^z_\alpha$ form factor is non-zero for $m''=m$ and therefore for a total number of rapidities in the intermediate states ($N''_p$) given $N''_p = N_p$. Using this fact, the selection rule on $J''$ and Eq. 
(\ref{corr}), we easily find that the only contributions come from states with 
$N''_r = \left\{N_r-1,N_r,N_r+1\right\}$.

Similarly, for  $S^-_\alpha$, we find that  $N''_p = N_p-1$ and the number of diverging rapidities must be given by $N''_r = \left\{N_r,N_r-1,N_r-2\right\}$.

For the specific case of ground state ($N_r=N_p$ diverging rapidities) expectation values, non-zero contributions to $S^z_\alpha$ correlations are found exclusively for intermediate states with either $N_p$ or $N_p-1$ diverging rapidities since $N''_r = N_p+1 > N_p$ is impossible. Identically, for $S^-_\alpha$, the only possible cases are given by an intermediate state with $N_p-1$ or $N_p-2$ diverging rapidities. 

Finally, for the $S^z_\alpha$ form factors involving only the ground state, one can use the fact that:

\bea
&&\lim_{g\to \infty} \left|GS\right> = \lim_{g\to \infty} \prod_{j=1}{N_p} \mathcal{C}(v_j) \left|0\right> \propto \left(S^+_{tot}\right)^{N_p}  \left|0\right> \nonumber\\ &=& \left|\uparrow_\alpha\right> \otimes  \sum_{\{ i_1, \ ... \ \alpha_{N_p-1}\}}^{\binom{N-1}{N_p-1}}  \left|\left\{\uparrow_{\alpha_1} ...\uparrow_{\alpha_{N_p-1}}\right\}\right> \nonumber\\ &&+ \left|\downarrow_\alpha\right> \otimes \sum_{\{ \alpha_1, \ ... \ \alpha_{N_p}\}}^{\binom{N-1}{N_p}} \left|\left\{\uparrow_{\alpha_1} ...\uparrow_{\alpha_{N_p}}\right\}\right>, 
\eea
 
 \noindent where   $\displaystyle \sum_{\{ \alpha_1, \ ... \ \alpha_{M}\}} \left|\left\{\uparrow_{\alpha_1} ...\uparrow_{\alpha_{M}}\right\}\right>$ is simply the sum over all possible states containing $M$ up spins picked out of the $N-1$ levels which exclude level $\alpha$. The action of $S^z_\alpha$ on this state is trivially given by multiplying by ${1}/{2}$ 
 while adding a minus sign to the second sum, which in the end allows us to simply prove that:
 
 \bea
&& \lim_{g\to \infty} \left<GS\right| S^z_\alpha\left|GS\right> \propto \frac{1}{2} \left[\binom{N-1}{N_p-1} - \binom{N-1}{N_p}\right] \nonumber\\ &&=\frac{1}{2} \frac{(N-1)!}{(N-N_p-1)!(N_p-1)!}  \left[\frac{2N_p - N}{(N-N_p)N_p}\right].
\label{gsgssz}
 \eea
  
 One immediately sees that the half-filling $2N_p = N$ case has the peculiar feature of having these `ground state to ground state' form factors go down to zero in the strong limit coupling.
  
\subsection{Correspondence between $g=0$ and $g \to \infty$ eigenstates} 
 \label{algo}
 \subsubsection{General algorithm}
 
An algorithm relating the $g=0$ structure of a state to the number of diverging rapidities it will have at $g \to \infty$ was already proposed in Ref. \cite{rsd-02,rsd-03}.  It necessitates the evaluation of various quantities for every possible partitions of the $N$ levels into 3 disjoint contiguous sets of levels. An equivalent result can be obtained through the simple following algorithm, which was discussed before in Ref. \onlinecite{fcc-09p}. 
 
\begin{figure}[t]
\includegraphics[width=7.5cm]{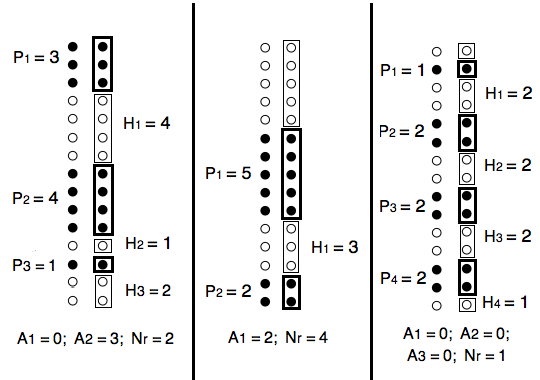} 
\caption{Construction of the contiguous blocks necessary to establish the $g=0$, $g\to\infty$ correspondence.}
\label{splitting}
\end{figure}

By splitting any $g=0$ configuration of rapidities into blocks of contiguous occupied and empty states, 
one can simply obtain the number of diverging roots. Fig. \ref{splitting} shows some examples of this construction. Every circle represents a single energy level and the blackened ones are occupied by a Cooper pair at $g=0$. 

We then label the various blocks according to the following prescription. The highest block of rapidities is labelled by index $i=1$ and contains $P_1$ rapidities. The block of unoccupied states right below it will be also labelled by $i=1$ and contains $H_1$ empty levels. We continue this labeling by defining $P_2 (H_2)$ as the number of rapidities (unoccupied states) in the next block until every single one of the $N_b$ blocks has been labelled. In the event that the lowest block ($i=N_b$) is a block of rapidities, as is the case in the middle example in Fig. \ref{splitting}, we set $H_{N_b}=0$.

The number of diverging rapidities is then simply given by:
\bea
 N_r =  \left[ P_{N_b}+A_{N_b-1} -\mathrm{Min}( P_{N_b}+A_{N_b-1},H_{N_b}) \right]
\eea

\noindent  with the $A_i$ terms defined recursively as:
\bea
A_i =  \left[ P_i+A_{i-1} -\mathrm{Min}(P_i+A_{i-1},H_i) \right]
\eea

\noindent with $A_0 = 0$.

\begin{figure}[b]
\includegraphics[width=2.8cm]{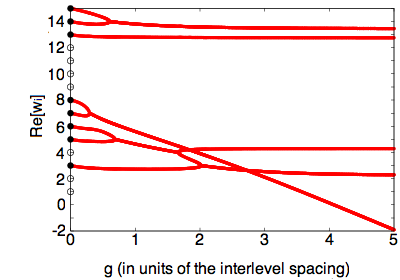} 
\includegraphics[width=2.8cm]{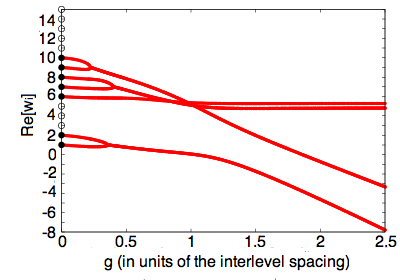} 
\includegraphics[width=2.8cm]{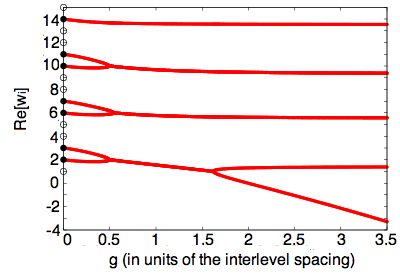} 
\caption{Evolution of the rapidities (real part) from 0 to large $g$ for the states presented in Fig. 1. }
\label{scan}
\end{figure}

This can be thought of as a `dynamical' process which is best understood by looking at the $g$ evolution of the rapidities as shown on Fig. \ref{scan} for the three states in Fig. \ref{splitting}.

As $g$ rises, each rapidity has a tendency to go down towards $-\infty$. Any block of $H_i$ unoccupied states stops up to $H_i$ rapidities from doing so by keeping them finite. If $P_i$ rapidities are going down and they meet a block of $H_i$ unoccupied states every rapidity will be kept finite if $H_i \geq  P_i$ ($A_i = 0$ rapidities will go through). On the other hand, whenever $H_i <  P_i$, only $H_i$ rapidities can be kept finite and the remaining $A_i = P_i-H_i$ will keep going down towards $-\infty$. Starting from the highest block of rapidities (of size $P_1$) we therefore have $A_1=P_1-\mathrm{Min}(P_1,H_1)$ which go through the $H_1$ empty states below. These will be added to the following block of $P_2$ rapidities giving $P_2+A_1$ rapidities which then meet a block of $H_2$ unoccupied states. $A_2=(P_2+A_1)-\mathrm{Min}(P_2+A_1,H_2)$ will go through and continue their descent. Keeping this analysis going until we reach the last blocks gives out the result given above. 

\subsubsection{Application to ground states and single excitation states}

\begin{figure}[h]
\includegraphics[width=7.5cm]{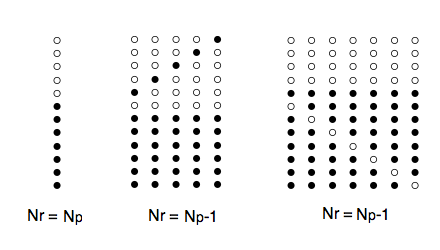} 
\caption{A ground state and the corresponding set of states which give a single finite rapidity at $g\to \infty.$}
\label{blocksa}
\end{figure}

At $g=0$, for any of  the canonical ground states containing $N_p$ rapidities in the $N_p$ lowest energy levels, we have a single block of unoccupied $N-N_p$ levels and a single block of $N_p$ occupied levels (see the left group in Fig. \ref{blocksa}). Since no unoccupied levels are below this $N_p$-rapidities block, they would all diverge in the $g \to \infty $ limit. 

Focusing on the states built out of a single excitation above this ground state, we find 3 distinct cases. When one moves the top rapidity (from level $N_p$) to a higher level (say level $\b > N_p$), the resulting block structure is $(P_1 =1)(H_1=\b-N_p) (P_2=N_p-1) $ (see middle group in Fig. \ref{blocksa}) which leads in the strong coupling limit to the single $P_1$ rapidity being kept finite while the remaining $N_p-1$ will diverge. In the second scenario, if one promotes any of the rapidities at level $\a < N_p$ (excluding the topmost one) to the level $N_p+1$ the resulting structure is $(P_1 =N_p-\a+1)(H_1=1) (P_2=\a-1)   $ (see the right group in Fig. \ref{blocksa}). Once again, this leads to one of the $P_1$ rapidities staying finite due to the unoccupied block $H_1=1$ while the other $N_p-1$ rapidities will diverge. 

The last possible case is a single excitation obtained by moving a rapidity from level $\a < N_p$ into an empty level $\b > N_p+1$. Doing this gives the following block structure $(P_1=1) (H_1= \b-N_p-1) (P_2 =N_p-\a) (H_2=1)(P_3=\a-1)$. In the end the $P_1$ rapidity will stay finite because of $H_1$, and one of the $P_2$ rapidities will remain finite as well, leading to a total of two finite rapidities.

From very simple combinatorics we therefore find that we can build $N-1$ single finite rapidity states (at $g\to \infty$) by deforming $g=0$ two-particle states. 

Moreover, at $g\to \infty$ for a given number of rapidities, the total number of states with $J=N_p-1$ (single finite rapidity) is given by the total number of solutions to the single Bethe equation $\sum_{i=1}^N \frac{1}{\lambda-\epsilon_i} = 0$ \ \cite{bortz-07}. This equation also has $N-1$ distinct solutions and therefore every state with a single finite rapidity at strong coupling stems from one of the two-particle states at $g=0$. This was pointed out before in Refs. \onlinecite{rsd-02,rsd-03}. 

 From the previous subsections we concluded that for the $S^z_\alpha$ form factors one can get contributions coming from the $N_p$ rapidities states with either $N_p$ or $N_p-1$ of them diverging. The first case we now know corresponds to the ground state.  Since we also showed that every state with one finite rapidity is generated by singly excited states, it becomes clear that the two-particle states do give out every non-zero contributions in the $g \to \infty$ limit. For $S^-_\alpha$ we showed that only the $N_p-1$ ground state (all rapidities divergent) or the $N_p-1$ states with $N_p-2$ diverging rapidities contribute. Once again this means that the intermediate sum can be limited to the $N_p-1$ ground state and single excitation (two-particle) states. One should also notice that, in this limit, any single excitation state which leads to two finite rapidities will not contribute although at weaker coupling they could.

Since the complete set of two particle states (plus the ground state) saturates the sum rules in both the $g \to 0$ and $g \to \infty$ limit, it is reasonable to assume that they will also be largely dominant in the crossover regime. This fact will be explicitly proven numerically since even for the smallest systems, this subset represents, for any $g$, more than 99\% of the weight.

\section{Dynamical correlation functions}
\label{seccorr}

We have presented all the ingredients to calculate the dynamical correlation functions: 
we need to solve the Richardson equations for each state $|\{ w\}\rangle$, calculate its energy $E_w$,
use the rapidities defining the solution to compute the determinant and calculate the form factors. 
However, while the formulas we obtained for the correlation functions are completely  
general and are valid for any choice of the Hamiltonian parameters $\e_\a$ and  $g$, 
to obtain a physical result we still have to perform the sum over the states and
this cannot be done analytically. 
Thus  we need to make a choice of the model to study.
As we already mentioned, we only consider the most-studied case in the
condensed matter literature, which consists of $N$ equidistant levels at
half-filling, i.e. $N = 2N_p$.  
We define the levels as
\be
\e_\a= \a\, \qquad {\rm with}\,\; \a=1\dots N\,,
\ee
i.e. we measure the energy scale in terms of the inter-level spacing and
we fix the Debye frequency (the largest energy level) to $N$.

\subsection{Diagonal $S_z$ correlator}

We start our analysis with the diagonal $S_z$ correlator that, in frequency space, reads
\be
G_{zz}^d(\omega)=\sum_{n=1}^N\sum_{\{ v\}} 
\frac{ |\langle \{ v\} |S^z_n|GS\rangle|^2}{
\langle GS| GS \rangle\langle \{ v\} | \{ v\} \rangle}
\delta(\omega-E_{ v}+E_{GS}) \,,
\label{gzzomega}
\ee
with $ | \{ v\} \rangle$ having $N/2$ rapidities.
In particles language this is a density-density correlator.
At any finite $N$, this is a sum of $\delta$ peaks each at the energy of the excited state $ | \{ v\} \rangle$, 
and each weighted with the corresponding form factor. 

In the thermodynamic limit where $N \to \infty$ while the inter-level spacing $d\to0$ (keeping a finite bandwidth for the single particle excitations), we can think of any $g > 0$ as being already the strong coupling case, i.e.: the BCS mean-field treatment becomes exact. In such a case correlation functions should be described by the $g\to \infty$ limit where only the single finite rapidity band would contribute. However, we are interested here in mesoscopic effects at finite $N$, that are encoded in the quantization of the energy levels and in the potential presence of non-trivial contribution from other excited states.

\begin{figure}[t]
\includegraphics[width=8.0cm]{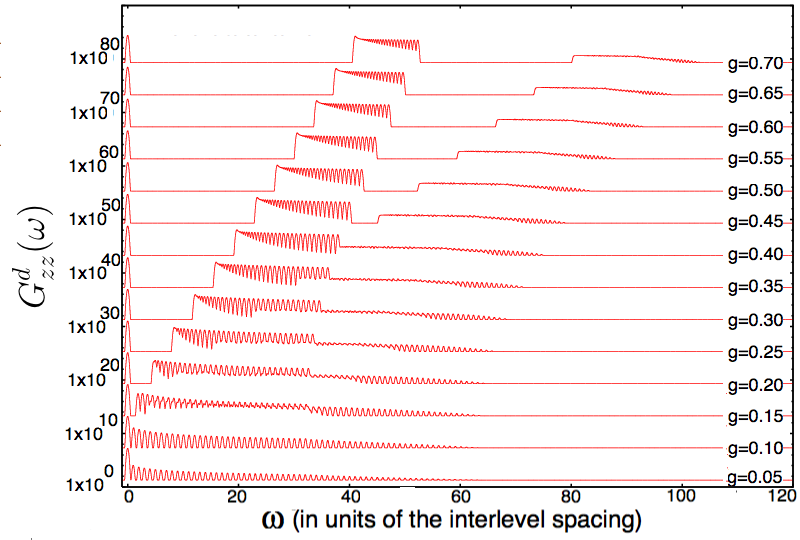} 
\caption{Diagonal correlation function $G_{zz}^d(\omega)$ obtained by smoothing the 
energy $\delta$-function with a Gaussian of width $w_G=0.1$ for different values of $g$ and at fixed number of pairs $N_p=32 \ (N=64)$.
We show $C(g)+G_{zz}^d(\omega) (10^6)^{\frac{g}{0.05}}$ with the offset $C(g) = \displaystyle \sum_{n=0}^{\frac{g}{0.05}-1} 20(10^6)^n$ (Effectively this makes scaled logscale plots with an offset). 
}
\label{freq-peak}
\end{figure}

\begin{figure}[t]
\includegraphics[width=8.0cm]{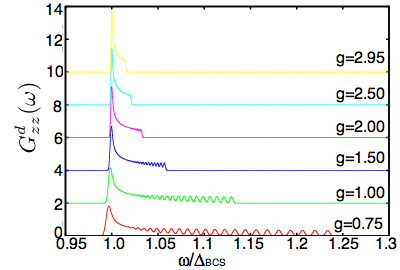} 
\caption{Diagonal correlation function $G_{zz}^d(\omega)$ obtained by smoothing the 
energy $\delta$-function with a Gaussian of width $w_G=0.1$ for different values of $g$ and at fixed number of pairs $N_p=32$.
We present the first excited subband contribution plotted with the frequency rescaled by the 
grand-canonical BCS gap, bringing the bottom of the band at ${\omega}/{\Delta_{BCS}} \approx 1$.}
\label{freq-peak2}
\end{figure}
 
In any real physical system, the $\delta$-peaks are smoothed by different effects
such as temperature broadening or, even in the case of very small $T$, experimental resolution.
All these effects are expected to broaden the $\delta$-peaks in an approximately Gaussian fashion. 
For this reason, in Fig. \ref{freq-peak} we plot a typical example of such correlator for different
values of $g$, at fixed number of pairs $N_p=32$, with the $\delta$ functions broadened to Gaussians of width $w_G=0.1$. The necessity to use logscale plots comes from the fact that typically the second energy subband, clearly separated by a gap for $g \ge 0.45$ has contributions which are orders of magnitude below the contributions from the first excited subband. In the limit $g\to \infty$, the width of these two subbands would go to zero while the gap separating the first subband from the ground state and the second band from the first, would tend to the grand-canonical BCS gap:

\be
\Delta_{GC}=\frac{N}{2\sinh (1/2g)}\,.
\label{Dgc}
\ee

The quantization of the energy levels is evident, especially for small $g$. 
For larger values of $g$, a peak seems to develop at the BCS gap even for these relatively small values of  $N$.  The contributions from the first excited subband are made more apparent in Fig. \ref{freq-peak2} which shows only this subband. We can understand this peak developing at the bottom of the band as being mostly due to the growing density of states at this energy. It is not due to a single large contribution from a given state, but simply to the finite width of the gaussian peaks making the small energy differences unresolvable. In fact, as shown explicitly in appendix \ref{scaling}, in the strong coupling limit all states in the subband have identical form factors and therefore equal contributions to the correlation function. The coalescence of the energies happening faster (in $g$) at the bottom of the band lead to this impression of a developing peak.

 Very similar plots can be found frequently in the experimental literature (see e.g. the review [\onlinecite{dr-01}]) for the I-V characteristic of superconducting nano-grains. 
We are now in a position to understand these results quantitatively.

\begin{figure}[b]
\includegraphics[width=8cm]{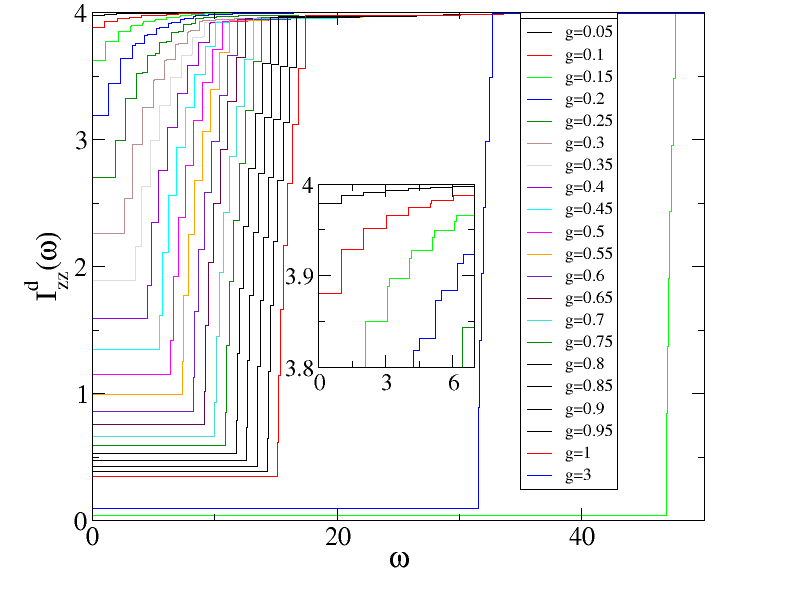} 
\caption{Integrated diagonal correlation function $I_{zz}^d(\omega)$ for $N=16$ and varying $g$.
The inset is a zoom of the upper left corner, where the quantization of the energy level is more evident.}
\label{szdiaglow}
\end{figure}

However, without a realistic idea of the kind and amplitude of broadening these plots are still only 
indicative. Moreover, it is difficult to correctly visualize these functions and the relative importance of the various contributions. A more precise information about the mesoscopic effects is encoded 
in the integrated correlation function obtained from $G^d_{zz}$ integrating it up to a given $\omega$
\be
I_{zz}^d(\omega)= \int_0^\omega d\omega' G_{zz}^d (\omega')\,,
\ee
i.e. the sum of the form factors of states with energy smaller than $\omega$.
The integrated correlation functions can be plotted without any smoothing. 
For several values of $g$, we report them in Fig. \ref{szdiaglow} for fixed $N=16$. 
Any step corresponds to a different eigenvalue (most of them have a two-fold degeneracy) of the Hamiltonian.
Notice that for large $\omega$ the integrated correlation function gives the value of the sum rule being 
 the sum of all form factors, i.e.  $I^d_{zz}(\omega=\infty)=\langle (S_{tot}^z)^2\rangle =N/4$ 
 (cf. Eq. (\ref{srzz})).

\begin{figure}[t]
\includegraphics[width=7.8cm]{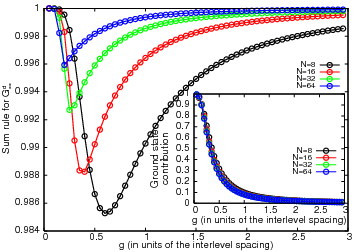}  
\caption{Sum-rule for the diagonal $G_{zz}^d(\omega)$ correlation function. 
Inset: The ground state contribution. Main Plot: Ground state plus all two-particle states.}
\label{szdiagrule}
\end{figure}

To understand how the various excitations combine to give this correlation function, in 
Fig. \ref{szdiagrule} we report  the contribution of the ground state and of the two-particle 
states to the sum rule. 
The ground state (inset) accounts for the full correlation function in the absence of interaction, 
(as discussed previously) but its contribution quickly decays to zero with increasing $g$, as
consequence of a complete re-organization of the ground state structure. 
In the main plot of Fig. \ref{szdiagrule}, we show the sum of the ground state plus all two-particle states. 
It is evident that in all considered cases, these states give basically the full correlation function, 
and at  most about $1\%$ is left to the other states. 
In particular for large $g$, the two-particle states account for the full correlation function (as shown in the 
previous sections). Notice also that with increasing $N$, the missing contribution becomes smaller.
This property simplifies enormously the computation of the correlation functions.
In fact, to have an effective description of the correlation functions, we do not need to sum 
over the total $\binom {N}{N/2}$ states, but only over the $N^2/4$ two-particles states.
For this reason, in the various figures, all correlation functions have been calculated by only 
considering two-particle states (for lower $N$, we checked by full sums that this `approximation' 
does not introduce any visible modification.)

 \begin{figure}[b]
\includegraphics[width=7.8cm]{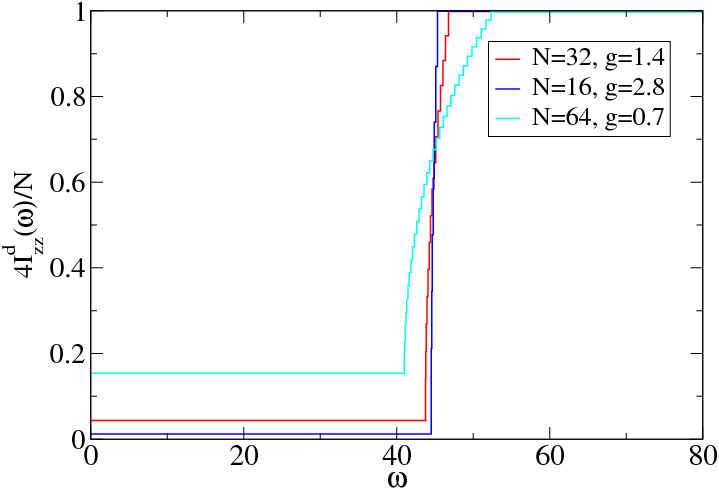}  
\caption{Integrated diagonal correlation function $I_{zz}^d(\omega)$ for  three values of $N$ and $g$,  
while keeping constant $Ng$.}
\label{szvarN}
\end{figure}

We can now come back to the analysis of the correlation function $I_{zz}^{d}(\omega)$ itself.
In Fig. \ref{szvarN} we plot three different curves at fixed $gN$ in such a way that the BCS gap is 
constant in the large $g$ approximation (see Eq. (\ref{Dgc}) giving the BCS gap in the grand-canonical ensemble). It is evident that for the smallest value of $N=16$, 
which corresponds to the largest value of $g=2.8$, $I_{zz}^d(\omega)$ has almost its asymptotic 
expression (see Appendix \ref{scaling}) that is a step function at the BCS gap (the integral of $\delta(\omega-\Delta)$ contribution for large $g$). Although the first energy band still has a finite width (which would go down to zero as $g^{-1}$ according to equation \ref{enerband}) this width is already significantly smaller than the BCS gap (which scales as $g$).
Oppositely for $N=64$ and $g=0.7$, despite $N$ being larger, the small value of $g$ and
the still significant relative width of the sub-band leads to a nice quantization of the energy spectrum, resulting in a staircase function with unequal steps. In the inset of Fig. \ref{szdiaglow}, the zoom of the lower energy sector for small $g$ at $N=16$ is reported, showing the formation of  intermediates steps of unequal contributions with increasing $g$.

\begin{figure}[t]
\includegraphics[width=7.8cm]{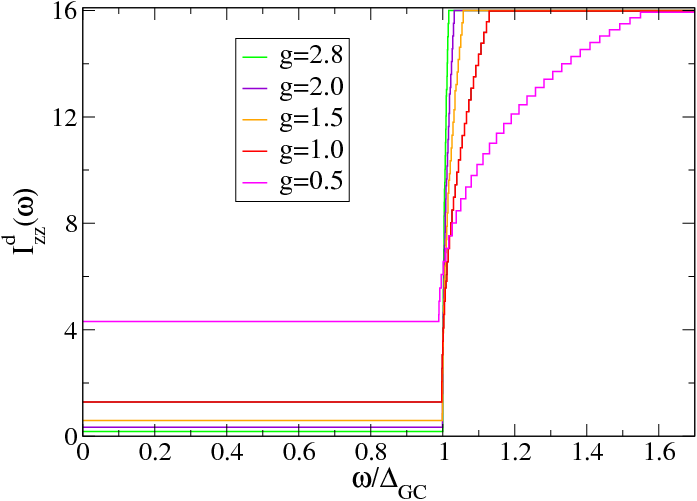}  
\caption{Integrated diagonal correlation function $I_{zz}^d(\omega)$ for $N=64$ and various $g$.
The horizontal axis has been rescaled by the BCS gap.}
\label{szdel}
\end{figure}

Finally in Fig. \ref{szdel}, we report $I_{zz}^d(\omega)$ at fixed $N=64$ and varying $g$, rescaling the 
horizontal axis by the grand-canonical BCS gap (eq. \ref{Dgc}).

For large $g$, $\Delta_{GC}$ corresponds exactly to the energy of the first excited state (and therefore sub-band) and all the curves start rising from $1$. However for smaller values of $g$ (in particular for $g=0.5$), visible finite size effects are present. As already stressed above, the staircase structure for small $g$, smoothly connects to a step function for large enough $g$.

\subsection{Diagonal $\langle S^+S^-\rangle$ correlator.}

\begin{figure}[t]
\includegraphics[width=7.8cm]{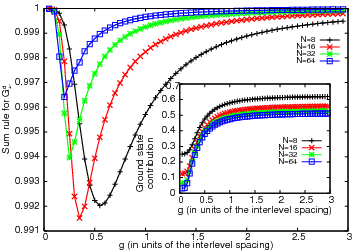}  
\caption{Sum-rule for the diagonal $G_{+-}^d(\omega)$ correlation function. 
Inset: The ground state contribution. Main plot: Ground state plus all two-particle states.}
\label{smdiagrule}
\end{figure}

This correlation function (as well as the corresponding off-diagonal one) 
is directly connected with the annihilation and creation of a Cooper pair.
We could imagine for example an experiment in which a superconducting grain was contacted on the left
and right by two separate bulk superconducting leads, each having their respective order parameters,
a problem which was studied for example in [\onlinecite{1993_Matveev_PRL_70,2007_Tanuma_PE_40}].  The Josephson
current in such a system is given by perturbation theory in the coupling between the leads and
the grain, and its calculation involves computing such correlators as the one above.  We will 
treat this problem in more detail in a separate publication, concentrating for the moment on
the results for the correlators themselves.

We start by showing the sum rule (c.f. Eq. (\ref{sr+-})) 
obtained from the ground state and all the two-particle states 
as function of $g$ at  fixed $N$ in Fig. \ref{smdiagrule}. 
For the ground state, we have an opposite behavior compared to the $S^z$ correlation:
the value for small $g$ is low and it increases by increasing $g$, but never saturating it completely.
The missing contribution is again mostly in the two-particle states, as evident from Fig. \ref{smdiagrule}.
For $g=0$ and $g\to\infty$, it has been proved analytically in the previous section that this class of states
gives a perfect saturation of the sum rule and hence the full correlation function.
The figure shows that even for intermediate $g$, a very accurate calculation comes only from 
two-particle states, and so in the following we will ignore all the other too small contributions. 
Notice that the contribution of the leading excitations is even 
more relevant than for $G_{zz}^d$, having always saturation above $99\%$ and that again for large 
enough $N$ it increases while increasing $N$.

\begin{figure}[b]
\includegraphics[width=9.0cm]{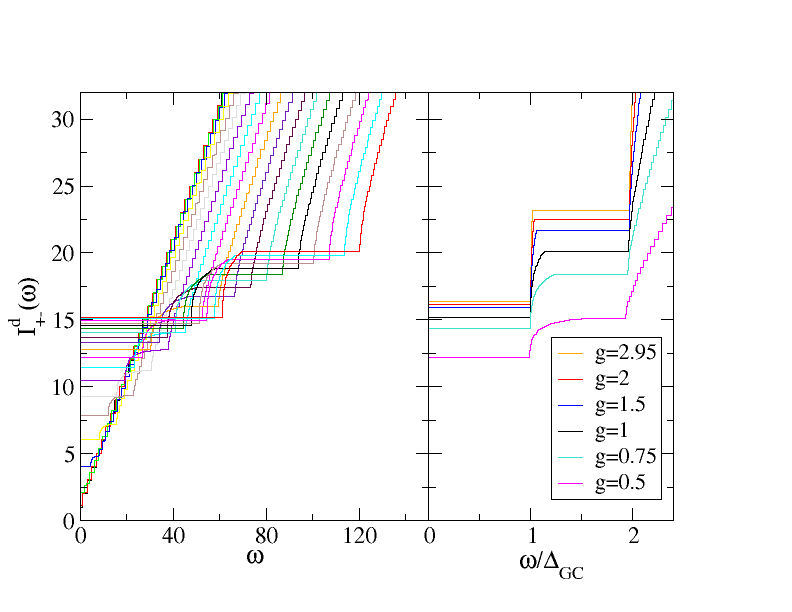} 
\caption{Integrated correlation function $I_{+-}(\omega)$ as function of $\omega$ for several couplings 
$g$ and fixed $N=64$. Left: small values of $g\leq1$. Right: as function of $\omega/\Delta_{GC}$ to show
the formation of two steps in the limit of large $g$.}
\label{smdiag}
\end{figure}

The integrated correlations
\be
I_{+-}^d(\omega)= \int_0^\omega d\omega' G_{+-}^d (\omega')\,,
\ee
 are reported in Fig. \ref{smdiag} for $N=64$.
In the left panel we report the small $g\leq1$. At very low $g$, the BCS gap is not yet  formed and the 
correlation function has a staircase structure with almost equal steps, as a consequence of the 
almost perfect equispacing of the energy level. Increasing $g$, the levels become incommensurate 
and the correlation function acquires a structure that reflects the formation of the BCS gap.
In the right panel we show the same correlation function plotted in terms of $\omega/\Delta_{GC}$ 
with the grand-canonical gap given by Eq. (\ref{Dgc}).
For large values of $g$, the `band structure' of the energy levels is evident in the 
formation of two main-steps at $\Delta_{GC}$  and at $2\Delta_{GC}$. 
In the same panel we also show some intermediate and small values of $g$ to show the formation of 
this two-step structure from the sharpening of the smaller steps. 
This is the main signature of mesoscopic effects in the correlation functions. Whereas for the $S^z$ correlations the second excited band's contribution were very rapidly supressed, in the case at hand, it maintains a very important contribution in the full regime studied in this paper. Although the selection rules point out that it no longer contributes at $g\to \infty$ it appears that the suppression of the $S^-$ form factors happens at much stronger coupling than it does for $S^z$. This seems to suggest that corrections to the mean field BCS results (more or less equivalent to the $g\to\infty$ limit) could show up in a wide regime when looking at the previously mentioned Josepshon current experiments.

\subsection{Off-diagonal correlators.}

\begin{figure}[t]
\includegraphics[width=7.8cm]{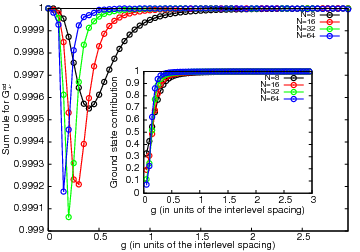}
\caption{Sum-rule for the global $G_{+-}^{od}(\omega)$ correlation function. 
Inset: The ground state contribution. Main plot: Ground state plus all two-particle states.}
\label{smrule}
\end{figure}

The off-diagonals correlators have properties very similar to the diagonal ones. For this reason we
only shortly present them, without an extensive discussion.
In Fig. \ref{smrule}, we report the contribution to the sum-rule (c.f. Eq. \ref{sr+-od})
from the ground state and from all two-particle states. 
The ground state by itself saturates the sum rule for large enough $g$, while for small values the contributions of the two-particle states are essential. Notice that the saturation of the sum-rule is always above $99.9\%$, even more than in the diagonal case. Since the ground state is the strongly dominant contribution to the correlation,  $G_{+-}^{od}$ will be a $\delta$ peak at the ground state value. 
Thus this is the less sensitive function to reveal mesoscopic effects.

\begin{figure}[b]
\includegraphics[width=8.5cm]{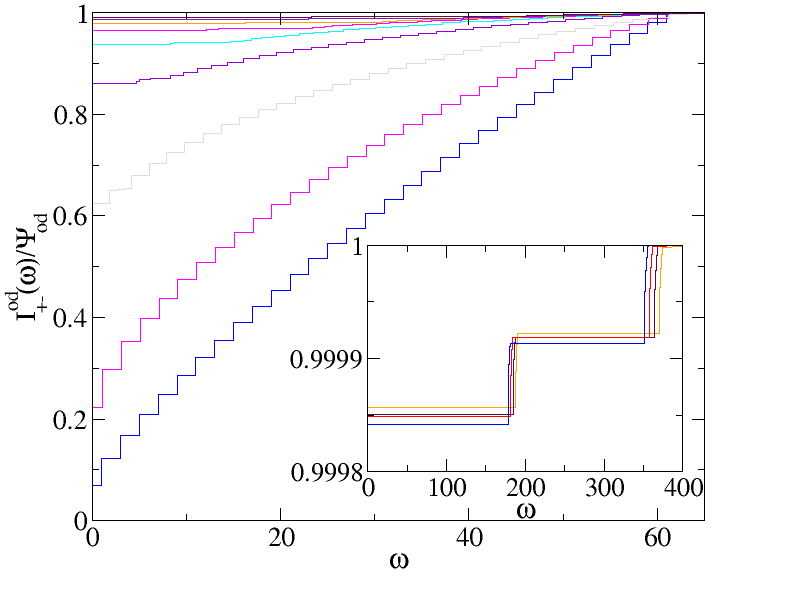} 
\caption{Integrated diagonal correlation function $I_{+-}^{od}(\omega)$. 
Main plot: Small values of $g\leq1$ at step of $0.05$ going from the staircase to the flat regime.
Inset: Large values of $g=2.8,2.85,2.9,2.95$ showing the very small two steps structure.}
\label{sm}
\end{figure}

In Fig. \ref{sm}, we report the integrated correlation function 
\be
I_{+-}^{od}(\omega)= \int_0^\omega d\omega' G_{+-}^{od} (\omega')\,,
\ee
for several $g$ and $N=64$. 
To make the plot visible we divide by the off-diagonal order parameter $\Psi_{od}$ as obtained by 
the Hellmann-Feynman theorem \cite{fcc-08}.
 In the main plot, small $g\leq1$  are reported. As before, for small enough $g$ we have 
a regular staircase behavior, that however becomes quickly flat because the ground state contribution 
to the correlation function is too large. For large values of $2.8\leq g< 3$ are reported in the inset of the 
figure and are zoomed very close to $1$. The contribution from two bands at $\Delta_{GC}$ and
$2\Delta_{GC}$ is evident, but it is visible only because we have an exact solution at hand. 
It would have been very hard, if not impossible, to see such a small effect in any numerical approach and
in a real experiment.

\begin{figure*}[t]
\includegraphics[width=7.2cm]{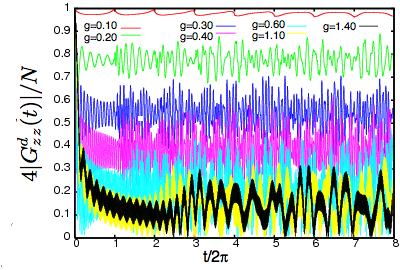} 
\includegraphics[width=7.2cm]{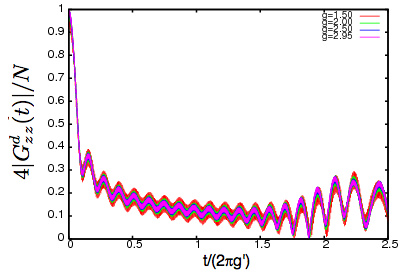} 
\includegraphics[width=7.2cm]{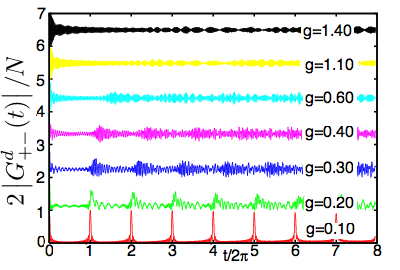} 
\includegraphics[width=7.2cm]{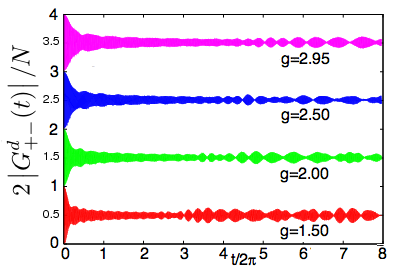} 
\caption{Real-time correlation function with varying $g$.
Top: $4|G_{zz}^{d}(t)|/N $; at small to intermediate coupling (left panel); as a function of rescaled time in the strong coupling regime (right panel) .
\\
  Bottom: $2|G_{+-}^d(t)|/N$; at small to intermediate coupling (left panel); at strong coupling (right panel). These functions have been shifted by an integer to make it easier to read, but they remain bounded between 0 and 1.}
\label{realt}
\end{figure*}

\subsection{Real time correlations.}

It is also interesting  to look at these correlation functions in real time. 
The presence of incommensurate energy levels gives indeed a quite complicated structure that greatly 
simplifies in the best known large and small $g$ behavior. 
We show in Fig. \ref{realt} the typical evolution of the two diagonal correlation functions. It is evident 
that for small $g$, the functions are almost periodic as a trivial consequence of the almost commensurate
levels (this is strictly true only at $g=0$, but at $g=0.1$ several periods should pass before the non-exact 
commensurability is manifest). Increasing $g$, the behavior becomes irregular and for very large $g$, new structure emerges due to the formation of a well defined energy subband structure.

As can be seen on the top right panel, for large enough $g$, we find a peculiar scaling behavior for the norm of the $G_{zz}^d$ correlation function. Indeed, dividing time by $g'$, the dimensionless coupling constant ($g'=g/d$ with $d$ the interlevel spacing) we find an almost perfect agreement between the curves for various strong coupling cases. The specific details allowing a clear understanding of this regime are presented in Appendix \ref{scaling} where we compute the $1/{g'}$ expansion of both the energies and the form factors. One should know, however, that looking independently at the real and imaginary parts of this correlation function they both still show a rapidly oscillating component associated with the gap frequency. It is only when looking at the norm of the correlation that this surprising scaling can be found. 

The same is not true for the $G_{+-}^d$ correlator shown on the bottom two panel of figure \ref{realt}. In this case, even when looking exclusively at the norm of the correlation a contribution oscillating at the very large energy gap frequency remains present.  Due to the time scales plotted, this rapid oscillation can barely be resolved on the figure for coupling strength larger than 0.40 but a rescaling of the plots clearly shows that they remain present.

\section{Conclusions}
\label{concl}

We have studied the dynamical correlation functions of the reduced BCS  
model in the canonical ensemble by means of Algebraic Bethe Ansatz techniques. 
We presented analytic selection rules in the weak and large coupling regimes.
For finite values of the coupling and for finite number of particles, the correlation functions are 
calculated numerically by summing over the form factors of the relevant states. 
We showed that two-particles states always gives saturation of the sum rules that is above $99\%$
for all $g$ and $N$ considered, in stark contrast with other integrable models like spin-chains and 
one-dimensional gases.
We presented and discussed extensively the crossover from small to large $g$, where quantum 
fluctuations have a dominant role and lead to a difference between the canonical and grand-canonical
ensemble, and where a perturbative treatment would provide incorrect results. 

Additionally, we showed unexpected  behavior of $S^z$ correlations at half-filling. The properties of these correlations give rise to a scaling law which is the exact opposite of what one would expect if the properties were dominated by the energy scale associated to the superconducting gap.

We remind the reader that while we only studied the case of $N$ non-degenerate 
equidistant energy levels at
half-filling (which is the most interesting model from the
condensed matter point of view \cite{dr-01,ds-99}), this is in no way a strict limitation of our approach,
which can be in principle applied to any choice of initial energy level distribution.
The description of pairing  in nuclei is for example a situation where other choices of the parameters $\e_\a$ are more
natural \cite{dh-03,bdh-04,zv-05,sv-07}.  These could be treated by a simple adaptation of our results.

The correlation functions we have provided have applicability in the phenomenology of
many experimental situations.  In further work, we will apply the current results to transport phenomena through
metallic nanograins coupled to normal and/or superconducting leads.

\section*{Acknowledgments}

All the authors are thankful for support from the Stichting voor
Fundamenteel Onderzoek der Materie (FOM) in the Netherlands.
PC thanks ESF (INSTANS activity) for financial support. AF's work was supported by the DFG through SFB631, SFB-TR12 and the Excellence Cluster "Nanosystems Initiative Munich (NIM)".

\appendix

\section{Solving Richardson equations}
\label{secgs}

As explained in our previous publications \cite{fcc-08,fcc-09,fcc-09p} (see also 
[\onlinecite{rsd-02,r-66,s-07,rnd-57,ded-06}]), 
we find the solutions starting from the trivial $g=0$ solutions and slowly raising the value of the coupling constant. With a linear regression using the previously found solutions, one gets an appropriate guess for the new solution at $g+\delta g$. Using a simple Newton algorithm for solving coupled non-linear algebraic equations is then sufficient, provided one makes the necessary change of variables. 

Indeed, at any value of the coupling, rapidities are either real or form complex conjugate pairs (CCP) and the pairing of two rapidities only occurs at bifurcation points $g^*$ at which they are both exactly worth $\epsilon_c$, i.e. one of the single particle energy levels. It is also possible, as $g$ is increased, that two paired rapidities split apart becoming both real again. At a critical $g^*$ where a pair splits or forms, the derivatives $\frac{dw_i}{d g}$ are not defined making the computation of the necessary Jacobian impossible. This problem is easily circumvented by, in the vicinity of critical point at which $w_i=w_j=\epsilon_c$, making the following change of variables 
\bea
\lambda_+ &=& w_i+w_j
\nonumber\\
\lambda_- &=& (w_i-w_j)^2.
\eea

 On both sides of $g^*$, those are two real variables and they have well defined derivatives even at the bifurcation point. The fact that $g$ is slowly increased allows us to figure out beforehand whether given rapidities are about to form (or break) complex conjugate pairs. Naturally, it makes the numerical procedure more tedious than it would be if one was able to guess correctly the structure at the precise value of $g$ in which we are interested. However the lack of known analytical results about the solutions to these precise Bethe equations forces us to use this scanning procedure. Fortunately, for the dynamical correlations we only need a very restricted set of states in order to get a very accurate description. Since single solutions are addressed one by one independently of the dimension of the full Hilbert space, the problem remains numerically tractable for fairly large system sizes which matrix diagonalization could not tackle.

\begin{figure}[h]
\includegraphics[width=4.25cm]{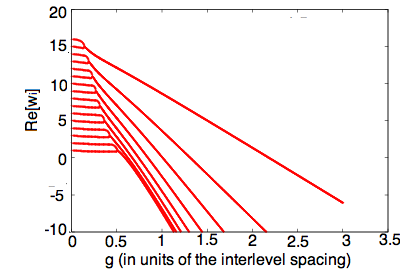}
\includegraphics[width=4.25cm]{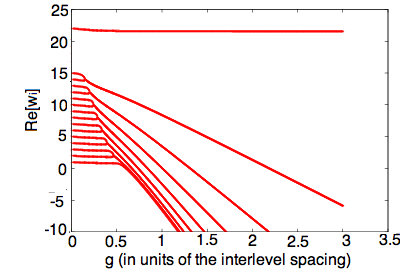}
\includegraphics[width=4.25cm]{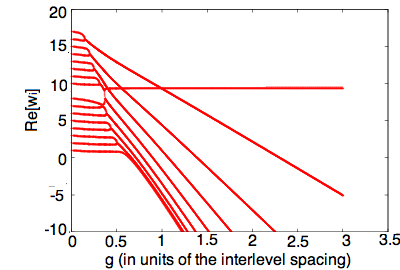}
\includegraphics[width=4.25cm]{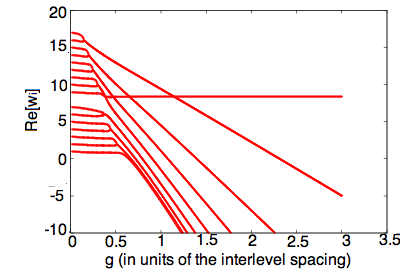}
\includegraphics[width=4.25cm]{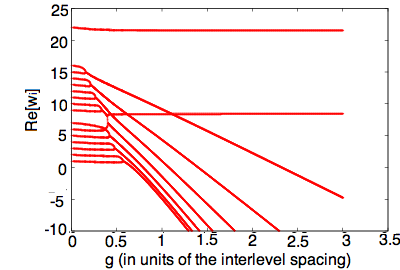}
\includegraphics[width=4.25cm]{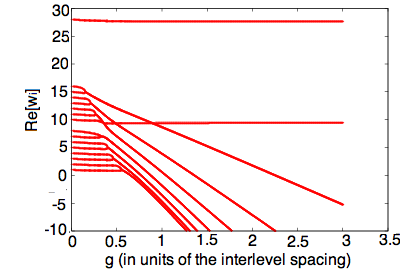}
\caption{Real part of rapidities for: ground state (upper left), promoted top ($\alpha=N_p$) rapidity (upper right), rapidity promoted at $\beta =N_p+1$ right above the Fermi level (middle left), rapidity promoted at $\beta =N_p+1$ right above the Fermi level (middle right), non-top rapidity promoted above $N_p+1$ level (lower left), non-top rapidity promoted above $N_p+1$ level (lower right)}
\label{figrap}
\end{figure}

In figure \ref{figrap}, we show the real part of the $N_p=16, N=32$ rapidities as a function of $g$ for typical single excitation states. As in the rest of the paper the lowest single particle energy level is chosen to be $\epsilon_1=1$ and the subsequent are at $\epsilon_\alpha=\alpha$.

For the ground state, rapidities form complex conjugate pairs in a very simple fashion, the top one pairing with the next one below and so on. For an odd number of rapidities, the lowest one would remain real and the other $N_p-1$ form CCPs in this way.

 In the top right pannel, we show a state obtained by promoting to level $\beta > N_p+1$ the highest rapidity (level index $\alpha = N_p$) from the ground state.  We know from the algorithm presented in section \ref{algo} that this state will have a single finite rapiditiy at strong coupling. Moreover, one sees that the promoted rapidity will simply stay real and remain between $\epsilon_{\beta}$ and $\epsilon_{\beta-1}$. Since there is no other rapidity close by, it cannot form a pair to go through the energy level $\epsilon_{\beta-1}$. Indeed, the structure of the Richardson equations prevents a single rapidity to be equal to an energy level $\epsilon_c$ since the diverging term $\frac{1}{w_i-\epsilon_c}$ has to be cancelled by a diverging term $\frac{-2}{w_i-w_j}$. 
 The remaining rapidities will simply form CCPs as they would for a $N_p-1$ pairs ground state.

The two figures in the middle are built by promoting a rapidity with index $\alpha < N_p$ to level $\beta=N_p+1$.  In both cases, we find only a single finite rapidity at strong coupling but we still have two different scenarios. The top contiguous block has an even (odd) number of rapidity on the left (right) pannel. For an odd number of rapidities in the top block, the lowest one from that block will simply stay real and between $\epsilon_{\alpha+1}$ and $\epsilon_\alpha$. In the even case, the lowest two rapidities from the top block will actually form a pair at the energy level $\alpha+1$, then split apart into two real rapidities at level $\alpha$. This splitting will make one rapidity be above $\epsilon_\alpha$ where it will stay until $g \to \infty$ while the other one goes below $\epsilon_\alpha$ and will later form a pair with the top rapidity from the block underneath.

Finally, the lowest pannels which lead to two finite rapidities at strong coupling behave in a similar way for the lowest $N_p-1$ rapidities, whereas the promoted one stays finite in the same way it did in the top right pannel. 

At $g\to\infty$ the state with one single finite rapidity in between $\epsilon_{\beta-1}$ and $\epsilon_\beta$, for $N_p \leq \beta \leq N$ is the deformed version of the $g=0$ state built by exciting the ground state's top rapidity $w_i= \epsilon_{N_p}$ to level $\epsilon_\beta$. On the other hand if the finite rapidity is between $1 < \alpha < N_p$ and $\alpha+1$ the state is the deformed version of $g=0$ state obtained by promoting the ground state's rapidity $w_i= \epsilon_\alpha$ to level $\epsilon_{N_p+1}$. The approximative location (between given energy states) of these strong coupling finite rapidities was already know and used in \cite{bortz-07} to compute non-equilibrium dynamics in the related central-spin model. However we here show how they each correspond to a known $g=0$ state deformed by interactions.

\section{Strong coupling scaling}
\label{scaling}

Hand-waving arguments tend to lead to the idea that the excitation gap should be the dominant energy scale in this system at least for strong coupling where one expects the BCS description to be more or less adequate. We would therefore expect correlations in time to show strong oscillations at a frequency given by the gap (i.e. $\propto g$ in the strong coupling limit). Surprisingly, the $S^zS^z$ correlations shown in figure \ref{realt} actually show the exact opposite behavior. The magnitude of correlations do show scalable behavior but they happened to be slowed down by an increasing $g$ and therefore an increasing gap.

This appendix aims at understanding this peculiar scaling property. In order to do so, we can rely on the strong coupling expansion of the Richardson equations. The following analysis will give us `semi-analytical' (i.e. getting the numerical values still needs numerical work) expressions for both the energies and the form factors needed to understand this correlator.

We assume a convergent expansion exists around the $g\to \infty$ point for the values of the rapidities themselves (Fig. \ref{figrap}  shows this assumption to be correct at strong enough $g$). Keeping only the first relevant corrections, we can write down, for the 3 types of states we are interested in:

\noindent the ground state

\bea
w_j = C^{N_p}_j g +A_j+B_j \frac{1}{g}\ \ \     \forall \ j = 1 ... N_p
\eea

\noindent the single finite rapidity states

\be
\begin{cases}\displaystyle
\eta_k = \eta^\infty_k + \delta_k \frac{1}{g}
\\ \displaystyle
w^k_j = C^{N_p-1}_j g +A^{k}_j+B^k_j \frac{1}{g}
\end{cases}
 \forall \ j = 1 ... N_p-1
\ee

\noindent and states with two finite rapidities

\be
\begin{cases}\displaystyle
\eta_{1,k} = \eta^\infty_{1,k}+ \delta_{1,k} \frac{1}{g}
\\ \displaystyle
\eta_{2,k} = \eta^\infty_{2,k}+ \delta_{2,k} \frac{1}{g}
\\ \displaystyle
w^k_j = C^{N_p-2}_j g +A'^{k}_j+B'^k_j \frac{1}{g}
\end{cases}
 \forall \ j = 1 ... N_p-2.
\ee

In the last two cases we have respectively either $N-1$ or $(N_p)^2-N_p N-N-1$  possible values of $k$ each associated with a different possible solution of the Richardson equations (different possible values of the $\eta$ rapidities which remain finite). In all three cases, eq. \ref{laguerre} gives the values of the $C$ constants which define the diverging rapidities in the $g\to \infty$ limit which are therefore independent of the choice of finite rapidities values for any given $k$. One should understand that $g$ here is considered to be the dimensionless quantity $\frac{g}{d}$, $d$ being the interlevel spacing.

\subsection{Energies}

The respective energies of these states, obtained by summing the $N_p$ rapidities are therefore simply given by:

\bea
E_{GS} = \left[\sum_{j=1}^{N_p}C^{N_p}_j\right] g + \left[\sum_{j=1}^{N_p}A_j\right] + \left[\sum_{j=1}^{N_p}B_j\right] \frac{1}{g}
\eea

\bea
E_{k} &=& \left[\sum_{j=1}^{N_p-1}C^{^{N_p}-1}_j\right] g + \left[\eta^\infty_k+\sum_{j=1}^{N_p-1}A^k_j\right] \nonumber\\&&+ \left[\delta_k + \sum_{j=1}^{N_p-1}B^k_j\right] \frac{1}{g}
\eea

\bea
E_{\eta_{1,k},\eta_{2,k}} &=& \left[\sum_{j=1}^{N_p-2}C^{{N_p}-2}_j\right] g + \left[\eta^\infty_{1,k}+\eta^\infty_{2,k}+\sum_{j=1}^{N_p-2}A'^k_j\right] \nonumber\\ &&+ \left[ \delta_{1,k}+ \delta_{2,k} + \sum_{j=1}^{N_p-2}B'^k_j\right] \frac{1}{g}.
\eea

The expansion of the Richardson equations around the strong coupling solutions with a single finite rapidity gives us:

\bea
\frac{-\omega_j}{g} &=& \sum_{\alpha=1}^N \frac{1}{1 - \frac{\epsilon_\alpha}{\omega_j}}  -2 \sum_{j'\ne j}^{N_p-1}\frac{\omega_j}{\omega_j -\omega_{j'}}   -2\frac{1}{1 - \frac{\eta_k}{\omega_j}}
\eea

\bea
-C_j^{N_p-1} -A_j^k\frac{1}{g}&&\approx (N-2)-2\sum_{j'\ne j}^{N_p-1}\frac{C_j^{N_p-1}}{C_j^{N_p-1}-C_{j'}^{N_p-1}} \nonumber\\&&+ \left( \left[ \sum_{\alpha=1}^N  \epsilon_\alpha\right] -2\eta^\infty_k\right)\left[\frac{1}{C_j^{N_p-1} g} \right]
\nonumber\\ && 
+2\sum_{j'\ne j}^{N_p-1}\frac{(C^{N_p-1}_{j'}A^k_{j}-A^k_{j'}C^{N_p-1}_{j})}{(C^{N_p-1}_{j}-C^{N_p-1}_{j'})^2}\frac{1}{g}.
\nonumber\\
\eea

Which, order by order gives:

\bea
-C_j^{N_p-1} &=& (N-2) -2\sum_{j'\ne j}^{N_p-1}\frac{C_j^{N_p-1}}{C_j^{N_p-1}-C_{j'}^{N_p-1}} 
\label{order0}
\\
 -A_j^k &=& \left( \left[ \sum_{\alpha=1}^N  \epsilon_\alpha\right] -2\eta^\infty_k\right) \frac{1}{C_j^{N_p-1} }\nonumber\\&&+2\sum_{j'\ne j}^{N_p-1}\frac{(C^{N_p-1}_{j'}A^k_{j}-A^k_{j'}C^{N_p-1}_{j})}{\left(C^{N_p-1}_{j}-C^{N_p-1}_{j'}\right)^2}
\label{order1}
\eea

These equations are defined for any of the $N_p-1$ values of index $j$. Summing up the $N_p-1$ equations \ref{order0} (divided by $C_j^{N_p-1}$), we find 

\bea
-(N_p-1) &=& (N-2) \sum_j \frac{1}{C^{N_p-1}_j},
\eea

\noindent while summing up the  equations \ref{order1} gives us:

\bea
\nonumber\\
-\sum_j A_j^k &=&  \left( \left[ \sum_{\alpha=1}^N  \epsilon_\alpha\right] -2\eta^\infty_k\right) \sum_j \frac{1}{C^{N_p-1}_j }\nonumber\\ \sum_j A_j^k&=&  \left( \left[ \sum_{\alpha=1}^N  \epsilon_\alpha\right] -2\eta^\infty_k\right) \frac{(N_p-1)}{(N-2) }.
\eea

Using this last expression, the energies of every single finite rapidity state can be written including the lowest correction in $\frac{1}{g}$ as:

\bea
E_{k} &=&\left[\sum_{j=1}^{N_p-1}C^{{N_p}-1}_j\right] g + \left(\eta^\infty_k\right)\frac{N-2N_p}{N-2} \nonumber\\ &&+ 
\sum_{j=1}^{N_p-1}\left[B^k_j\right] \frac{1}{g} + \mathcal{O}\left(\frac{1}{g^2}\right).
\label{enerband}
\eea

This shows that the half-filled case leads to the complete energy collapse of the first excited band as proven before in [\onlinecite{yba-03}], i.e. $\displaystyle\lim_{g\to \infty}E_{k}-E_{k'} = 0 $. The zero bandwidth obtained in this specific case will be shown to be one of the central elements in the scaling properties of the $S^z$ operators dynamical correlations. 

\subsection{Eigenstates}

In order to establish a similar expansion for the states themselves, one can simply use their representation as a Bethe state (eq. \ref{RICHWF}) and expand the $\mathcal{C}$ operators used to construct them. Keeping terms up to order $\frac{1}{g}$ for both divergent $(w = Cg+A+\frac{B}{g})$ and finite ($\eta=\eta^\infty +\frac{\delta}{g})$ rapidities we have:

\bea
\mathcal{C}(w) &\approx& \sum_{\alpha=1}^{N} \frac{S^+_\alpha}{w-\epsilon_\alpha} =  \frac{1}{C g}\sum_{\alpha=1}^{N} S^+_\alpha \left[1+ \frac{\epsilon_\alpha-A}{C} \frac{1}{g}\right]
\nonumber\\
&=& 
  \frac{1}{C g}\left[S^+_{tot} -\frac{1}{g}  \frac{A}{C} S^+_{tot} +\frac{1}{g}  \frac{1}{C}\sum_{\alpha=1}^{N} \epsilon_\alpha S^+_\alpha \right]
\nonumber\\  
\mathcal{C}(\eta) &\approx&  \sum_{\alpha=1}^{N} S^+_\alpha \left[ \frac{1}{\eta^\infty-\epsilon_\alpha} - \frac{1}{g} \frac{\delta}{(\eta^\infty-\epsilon_\alpha)^2}\right].
\eea

By getting rid of the $\frac{1}{Cg}$ prefactors which, for physical quantities, will always be cancelled by equivalent factors in the norms, we can therefore write:

\bea
\left|GS\right>  &\approx& \left|GS^{N_p}\right> +\frac{1}{g} \sum_{\alpha=1}^N G_\a S^+_\a \left|GS^{N_p-1}\right> 
\eea

\bea
\left|\eta_k\right>  &\approx&\sum_{\a=1}^N F^k_\a S^+_\a \left|GS^{N_p-1}\right> \nonumber\\&& +\frac{1}{g} \sum_{\a,\b=1}^N G^k_{\a,\b} S^+_\a S^+_\b \left|GS^{N_p-2}\right> 
\eea

\bea
\left|\eta_{1,k},\eta_{2,k}\right>  &\approx&\sum_{\a,\b=1}^N F^k_{\a,\b} S^+_\a S^+_\b \left|GS^{N_p-2}\right> \nonumber\\
&&+\frac{1}{g} \sum_{\a,\b,\g =1}^N G^k_{\a,\b,\g} S^+_\a S^+_\b S^+_\g \left|GS^{N_p-3}\right> \nonumber\\
\eea

\noindent where the states $\left|GS^{M}\right> \equiv \left(S^+_{tot}\right)^M \left|0\right>$. The following set of definitions was also used:

\bea
G_\alpha&\equiv&  \left[\sum_{j=1}^{N_p} \frac{A_j}{C^{N_p}_j} \right] +  \left[\sum_{j=1}^{N_p} \frac{1}{C^{N_p}_j} \right] \epsilon_\alpha
\nonumber\\
F^k_\alpha&\equiv& \frac{1}{\eta_k^\infty-\epsilon_\alpha}
\nonumber\\
G^k_{\a,\b}&\equiv&
-\left[\sum_{j'=1}^{N_p-1} \frac{A^k_{j'}}{C^{N_p-1}_{j'}} \right] \frac{1}{\eta_k^\infty-\epsilon_\a}-\frac{\delta_k}{(\eta_k^\infty-\epsilon_\a)^2}
\nonumber\\ && + \left[\sum_{j'=1}^{N_p-1} \frac{1}{C^{N_p-1}_{j'}} \right] \frac{\epsilon_\b}{(\eta_k^\infty-\epsilon_\a)^2}
\nonumber\\
F^{k}_{\a,\b}&\equiv&  \frac{1}{\eta_{1,k}^\infty-\epsilon_\a} \frac{1}{\eta_{2,k}^\infty-\epsilon_\b}
\nonumber\\
G^k_{\a,\b,\g}&\equiv&
-\frac{\delta_{2,k}}{(\eta_{2,k}^\infty-\epsilon_\b)^2}\frac{1}{\eta_{1,k}^\infty-\epsilon_\a}
-\frac{\delta_{1,k}}{(\eta_{1,k}^\infty-\epsilon_\a)^2}\frac{1}{\eta_{2,k}^\infty-\epsilon_\b}
\nonumber\\&&
-\left[\sum_{j'=1}^{N_p-2} \frac{A^k_{j'}}{C^{N_p-2}_{j'}} \right] \frac{1}{\eta_{1,k}^\infty-\epsilon_\a}\frac{1}{\eta_{2,k}^\infty-\epsilon_\b}
\nonumber\\&&
+ \left[\sum_{j'=1}^{N_p-2} \frac{1}{C^{N_p-2}_{j'}} \right] \frac{\epsilon_\g}{(\eta_{1,k}^\infty-\epsilon_\a)(\eta_{2,k}^\infty-\epsilon_\b)}.
\eea

\subsection{Form factors}

At $g\to\infty$ it is straightforward to compute value of the various forms factors. For the ground state it was done previously (eq. \ref{gsgssz}). Although the first order correction can also be obtained in a similar fashion, we will not explicitly need the coefficients of the expansion and therefore simply write:

\bea
\left<GS\right|S^z_\alpha\left|GS\right>\approx
\frac{1}{2} \left[\binom{N-1}{N_p-1} - \binom{N-1}{N_p} \right] + \frac{1}{g} A_\alpha,
\nonumber\\
\eea

\noindent with  $A_\alpha$ an unspecified (although obtainable) constant.

Identically, for single finite rapidity states we can write:

\bea
&&\left<GS\right|S^z_\a\left|\eta_k\right>_{g\to \infty} =\sum_\b F^k_\b \left<GS^{N_p}\right| S^z_\a S^+_\b \left|GS^{N_p-1}\right>
\nonumber\\
&&= \sum_{\b \ne \a} F^k_\b \left<GS^{N_p}\right|  \nonumber\\ && \left[\frac{1}{2} \left|\uparrow_\b,\uparrow_\a \right> \otimes \sum_{\{ \a_1, \ ... \ \a_{N_p-2}\}}^{\binom{N-2}{N_p-2}} \left|\left\{\uparrow_{\a_1} ...\uparrow_{\a_{N_p-2}}\right\}\right>\right.
\nonumber\\ &&\left.-\frac{1}{2} \left|\uparrow_\b,\downarrow_\a\right> \otimes \sum_{\{ \a_1, \ ... \ \a_{N_p-1}\}}^{\binom{N-2}{N_p-1}} \left|\left\{\uparrow_{\a_1} ...\uparrow_{\a_{N_p-1}}\right\}\right>\right]
\nonumber\\&&+ F^k_\a \frac{1}{2}  \left<GS^{N_p}\right| \left[\left|\uparrow_\a\right> \otimes \sum_{\{ \a_1, \ ... \ \a_{N_p-1}\}}^{\binom{N-1}{N_p-1}} \left|\left\{\uparrow_{\a_1} ...\uparrow_{\a_{N_p-1}}\right\}\right>\right]
\nonumber\\
&&= \frac{1}{2} \sum_{\b\ne \a} F^k_\b \left[\binom{N-2}{N_p-2}-\binom{N-2}{N_p-1}\right] + \frac{1}{2}F^k_\a\binom{N-1}{N_p-1}
\nonumber\\ &&=
 \frac{1}{2} \sum_{\b} F^k_\b \left[\binom{N-2}{N_p-2}-\binom{N-2}{N_p-1}\right] \nonumber\\ &&+
 \frac{1}{2}F^k_\a\left[\binom{N-1}{N_p-1} - \binom{N-2}{N_p-2}+\binom{N-2}{N_p-1}\right]
\eea

The orthogonality of these states with the ground state also allows us to write:

\bea
&& \left<GS\right.\left|\eta_k\right>_{g\to \infty} \nonumber\\&& = \sum_{\b} F^k_\b \left<GS^{N_p}\right| \left[\left|\uparrow_\b\right> \otimes \sum_{\{ \a_1, \ ... \ \a_{N_p-1}\}}^{\binom{N-1}{N_p-1}} \left|\left\{\uparrow_{\a_1} ...\uparrow_{\a_{N_p-1}}\right\}\right>\right] \nonumber\\ && =
 \binom{N-1}{N_p-1} \left[\sum_{\b} F^k_\b\right] = 0,
\eea

\noindent and therefore, adding the next order term through an unspecified constant:

\bea
\left<GS\right|S^z_\a\left|\eta_k\right>&\approx& \frac{F^k_\a}{2}\left[\binom{N-1}{N_p-1} - \binom{N-2}{N_p-2}+\binom{N-2}{N_p-1}\right] \nonumber\\&& + \frac{B^k_\a}{g}
\nonumber\\ &=&  F^k_\a \frac{(N-2)!}{(N_p-1)!(N-N_p-1)!}+ \frac{B^k_\a}{g}. 
\eea

It was also proven in section \ref{strongg} that at $g\to\infty$ we have $\left<GS\right|S^z_n\left|\eta_{1,k},\eta_{2,k}\right>_{g\to \infty} =0$ and we therefore have:

\bea
\left<GS\right|S^z_\a\left|\eta_{1,k},\eta_{2,k}\right> \approx \frac{C_\a^k}{g}.
\eea

Finally one can similarly compute the squared norms of the ground state and the single rapidity states.

\bea
\left<GS\right.\left|GS\right>_{g\to \infty} &=&   \binom{N}{N_p}
\eea
\bea
\left<\eta_k\right.\left|\eta_k\right>_{g\to \infty} &=& \sum_{\a,\b} F_\a^k(F_\b^k)^*\left<GS^{N_p-1}\right|S^-_\b S^+_\a\left|GS^{N_p-1}\right>
\nonumber\\ &=& \sum_{\a\ne \b} F_\a^k(F_\b^k)^*\binom{N-2}{N_p-2} \nonumber\\&& +\sum_{\a} |F_\a^k|^2 \binom{N-1}{N_p-1}
\nonumber\\ &=& \sum_{\a,\b} F_\a^k(F_\b^k)^*\binom{N-2}{N_p-2} \nonumber\\&& +\sum_{\a} |F_\a^k|^2 \left[\binom{N-1}{N_p-1} - \binom{N-2}{N_p-2} \right]\nonumber\\ &=& 0+ \sum_\a |F_\a^k|^2  \frac{(N-2)!}{(N_p-1)!(N-N_p-1)!}.
\nonumber\\
\eea

Specializing to the half-filled case, we find:

\bea
&&\sum_{\a=1}^N\frac{\left|\left<GS\right|S^z_\a\left|GS\right> \right|^2}{\left<GS\right.\left|GS\right>\left<GS\right.\left|GS\right>}\approx \frac{A}{g^2}
\\
&&\sum_{\a=1}^N\frac{\left|\left<GS\right|S^z_\a\left|\eta_k\right> \right|^2}{\left<GS\right.\left|GS\right>\left<\eta_k\right.\left|\eta_k\right>}\approx 
 \left(\frac{N}{4(N-1)}\right)\nonumber\\ &&+\sum_{\a=1}^N\frac{2\mathrm{Re}\left[F_\a^k B_\a^k\right]}{g\sum_\b |F_\b^k|^2} \frac{(N/2)!(N/2)!}{(N)!} +\frac{B}{g^2}
 \\
&&\sum_{\a=1}^N\frac{\left|\left<GS\right|S^z_\a\left|\eta_{1,k},\eta_{2,k}\right>\right|^2}{\left<GS\right.\left|GS\right>\left<\eta_{1,k},\eta_{2,k}\right.\left|\eta_{1,k},\eta_{2,k}\right>} \approx \frac{C}{g^2}.
\eea

At order 0 in $\frac{1}{g}$ we therefore find that the only $N-1$ non-zero contributions coming from the form factors (the ones involving the single finite rapidity states) are actually all equal. At the next leading order the only contributions also come from the same reduced set of states. 

Using equation \ref{gzzomega} we can write the correlation function by summing over the $N-1$ possible values of $\eta_k$. At order $\frac{1}{g}$, we have

\bea
&& G_{zz}^d(\omega) \approx \sum^{N-1}_{k=1}\delta(\omega-E_k+E_{GS}) \left[
 \left(\frac{N}{4(N-1)}\right)\right.\nonumber\\ &&\left.+\sum_{n=1}^N\frac{2\mathrm{Re}\left[F_n^k B_n^k\right]}{g\sum_i |F_i^k|^2} \frac{(N/2)!(N/2)!}{(N)!} \right],
\eea

\noindent whose Fourier transform gives us

\bea
&&G_{zz}^d(t) \approx \sum^{N-1}_{k=1}e^{-i(E_k-E_{GS})t} \left[
 \left(\frac{N}{4(N-1)}\right)\right.\nonumber\\ &&\left.+\sum_{\a=1}^N\frac{2\mathrm{Re}\left[F_\a^k B_\a^k\right]}{g\sum_\b |F_\b^k|^2} \frac{(N/2)!(N/2)!}{(N)!} \right].
\eea

Looking exclusively at the magnitude of the correlations and therefore at phase independent properties of this correlator we have

\bea
&&\left|G_{zz}^d(t)\right|^2 \approx 
\sum^{N-1}_{k,k'=1}e^{-i(E_k-E_{k'})t} \left[ \left(\frac{N}{4(N-1)}\right)^2\right.\nonumber\\ &+&  \left.
\left(\frac{[(N/2)!]^2}{2(N-1)(N-1)!}\right)\frac{\mathrm{Re}\displaystyle\sum_{\a=1}^N\left[F_\a^k B_\a^k+F_\a^{k'} B_\a^{k'}\right]}{g\sum_\b |F_\b^k|^2}\right]\nonumber\\
\eea

Since equation \ref{enerband} tells us that 

\bea
E_k-E_{k'} \equiv \frac{\Delta_{k,k'}}{g},
\eea

\noindent we can write 

\bea
&&\left|G_{zz}^d(t)\right|^2 \approx 
\sum^{N-1}_{k,k'=1}e^{-i\Delta_{k,k'}(\frac{t}{g})} \left[ \left(\frac{N}{4(N-1)}\right)^2\right.\nonumber\\ &+&  \left.
\left(\frac{[(N/2)!]^2}{2(N-1)(N-1)!}\right)\frac{\mathrm{Re}\displaystyle\sum_{n=1}^N\left[F_n^k B_n^k+F_n^{k'} B_n^{k'}\right]}{g\sum_i |F_i^k|^2}\right].\nonumber\\
\eea

This shows that at strong enough coupling the dominant term

\bea
 \left(\frac{N}{4(N-1)}\right)^2 \sum^{N-1}_{k,k'=1}e^{-i\Delta_{k,k'}(\frac{t}{g})} 
\eea

\noindent is purely a function of $\frac{t}{g}$. This fact is only true at half filling where the ground state average of any $S^z_\a$ is zero. Moreover, half filling also allows fulfillment of the second necessary condition, the vanishing of width of the first excited band. The first effect makes the energy scale associated to the BCS gap irrelevant since only the first band of excited states is contributing to the correlations. The vanishing width of this band is then responsible for the `inverted scaling' by making the relevant energy differences smaller as $g$ increases.

This very peculiar half-filling scaling property makes it possible to slow down specific dynamical processes in this system by making the interaction stronger. Since they are only related to the spectrum and the condition $\left<GS\right|S^z_\a\left|GS\right> =0$, this scaling law would hold at half-filling for any possible correlations of $S^z_\a$ operators; be they local, global, intra or inter-level correlations. The magnitude of any one of the possible correlations would still follow a similarly scalable time evolution. 

Although clearly valid in the strong $g$ limit where $\frac{1}{g} << 1$, we also find through the numerical work carried out in this paper that this scaling behavior extends to a very broad range of coupling constants. For $g\gtrsim  1.5 d$ with $d$ the inter-level spacing, we find that $\frac{1}{g}$ corrections are already strongly suppressed and the scaling behavior is therefore already apparent.


\begin{thebibliography}{999}


\bibitem{bcs-57}
J. Bardeen, L. N. Cooper, and J. R. Schrieffer,
Phys. Rev. {\bf 106}, 162 (1957);  {\it ibid.} {\bf 108}, 1175 (1957).

\bibitem{exp}
D. C. Ralph, C. T. Black, and M. Tinkham,
Phys. Rev. Lett. {\bf 74}, 3241 (1995);
and ibid. {\bf 76}, 688 (1996);
and ibid. {\bf 78}, 4087 (1997).

\bibitem{dr-01}
J. von Delft and D. C. Ralph, Phys. Rep. {\bf 345}, 61 (2001).

\bibitem{rs-62}
R. W. Richardson, Phys. Lett. {\bf 3}, 277 (1963); {\bf 5}, 82 (1963);
R. W. Richardson and N. Sherman, Nucl. Phys. {\bf 52}, 221 (1964);
{\bf 52}, 253 (1964).


\bibitem{dps-04} J. Dukelsky, S. Pittel, and G. Sierra,
Rev. Mod. Phys. {\bf 76},  643 (2004)

\bibitem{dh-03} D. J. Dean and M. Hjorth-Jensen,
Rev. Mod. Phys. {\bf 75}, 607 (2003)

\bibitem{r-65}
R. W. Richardson, J. Math. Phys. {\bf 6}, 1034 (1965).

\bibitem{ao-02}
L. Amico and A. Osterloh, Phys. Rev. Lett. {\bf 88}, 127003 (2002).


\bibitem{zlmg-02}
H.-Q. Zhou, J. Links, R.H. McKenzie, and M.D. Gould,
Phys. Rev. B {\bf 65}, 060502(R) (2002).

\bibitem{lzmg-03}
J. Links, H.-Q. Zhou, R.H. McKenzie, and M.D. Gould,
J. Phys. A {\bf 36}, R63 (2003).

\bibitem{s-89}
N. A. Slavnov, Teor. Mat. Fiz. {\bf 79}, 232 (1989).


\bibitem{fcc-08}
A. Faribault, P. Calabrese, and J.-S. Caux,  
Phys. Rev. B {\bf 77}, 064503 (2008).

\bibitem{mff-98}
A. Mastellone, G. Falci, and R. Fazio, Phys. Rev. Lett. {\bf 80}, 4542 (1998).

\bibitem{sbb-08}
S. Staudenmayer, W. Belzig, and C. Bruder
Phys. Rev. A {\bf 77}, 013612 (2008).

\bibitem{sc}
J.-S. Caux and J. M. Maillet, Phys. Rev. Lett. {\bf 95}, 077201 (2005)
J.-S. Caux, R. Hagemans, and J. M. Maillet, J. Stat. Mech. P09003  
(2005).

\bibitem{afc-09}
V. Alba, M. Fagotti, and P. Calabrese,
J. Stat. Mech. (2009) P10020. 

\bibitem{1dbg}
J.-S. Caux and P. Calabrese, Phys. Rev. A {\bf 74}, 031605R (2006);
J.-S. Caux, P. Calabrese, and N. A. Slavnov, J. Stat. Mech. P01008  
(2007).



\bibitem{c-09}
J.-S. Caux, J. Math. Phys. {\bf 50}, 095214 (2009)

\bibitem{fcc-09}
A. Faribault, P. Calabrese, and J.-S. Caux,  
J. Stat. Mech. (2009) P03018.

\bibitem{fcc-09p}
A. Faribault, P. Calabrese, and J.-S. Caux, 
J. Math. Phys. {\bf 50}, 095212 (2009).

\bibitem{g-book}
M. Gaudin,
{\it Mod\`eles Exactement R\'esolus} (Les \'Editions de Physique,
Les Ulis, France, 1995).

\bibitem{crs-97} M. C. Cambiaggio, A. M. F. Rivas, and M. Saraceno,
Nucl. Phys. A {\bf 624}, 157 (1997).

\bibitem{aff-01} L. Amico, G. Falci, and R. Fazio,
J. Phys. A {\bf 34} 6425, (2001).

\bibitem{dp-02}
J. von Delft and R. Poghossian, Phys. Rev. B {\bf 66}, 134502 (2002).

\bibitem{dzgt-96}
J. von Delft, A. D. Zaikin, D. S. Golubev, and W. Tichy, Phys. Rev. Lett. {\bf 77}, 3189 (1996).

\bibitem{sild-01}
M. Schechter, Y. Imry, Y. Levinson, and J. von Delft, Phys. Rev. B {\bf 63}, 214518 (2001).

\bibitem{yba-03}
E. A. Yuzbashyan, A. A. Baytin, and B. L. Altshuler,
Phys. Rev. B {\bf 68}, 214509 (2003).

\bibitem{yba-05}
E. A. Yuzbashyan, A. A. Baytin, and B. L. Altshuler,
Phys. Rev. B {\bf 71}, 094505 (2005).

\bibitem{sg-06}
I. Snyman and H. B. Geyer, Phys. Rev. B {\bf 73}, 144516 (2006).

\bibitem{rsd-02}
J.M. Roman, G. Sierra, and J. Dukelsky, Nucl.Phys. B {\bf 634}, 483 (2002).

\bibitem{rsd-03}
J. M. Roman, G. Sierra, and J. Dukelsky, Phys. Rev. B {\bf 67} 064510 (2003).

\bibitem{bortz-07} M. Bortz and J. Stolze, Phys. Rev. B {\bf 76}, 014304 (2007).




\bibitem{1993_Matveev_PRL_70} K. A. Matveev, M. Gisself{\"a}lt, L. I. Glazman, M. Jonson and R. I. Shekhter, Phys. Rev. Lett. 70, 2940 (1993).
\bibitem{2007_Tanuma_PE_40} Y. Tanuma, Y. Tanaka and K. Kusakabe, Physica E 40, 257 (2007).

\bibitem{ds-99}
J. Dukelsky and G. Sierra, Phys. Rev. Lett. {\bf 83}, 172 (1999);
G. Sierra, J. Dukelsky, G. G. Dussel, J. von Delft, F. Braun, Phys. Rev. B {\bf 61}, 11890 (2000).

\bibitem{bdh-04}
A. Belic, D.J. Dean, and M. Hjorth-Jensen,
Nucl. Phys. A {\bf 731}, 381 (2004).

\bibitem{zv-05}
V. Zelevinsky and A. Volya, Nucl. Phys. A {\bf 752}, 325 (2005).

\bibitem{sv-07}
T. Sumaryada and A. Voyla, Phys. Rev. C {\bf 76}, 024319 (2007).

\bibitem{r-66}
R. W. Richardson, Phys. Rev. {\bf 141}, 949 (1966).

\bibitem{s-07}
M. Sambataro, Phys. Rev. C {\bf 75}, 054314 (2007).

\bibitem{rnd-57}
S. Rombouts, D. Van Neck and J. Dukelsky, Phys. Rev. C {\bf 69}, 061303  (2004).

\bibitem{ded-06}
F. Dominguez, C. Esebbag, and J. Dukelsky, J. Phys. A {\bf 39}, 11349  (2006).








\end{thebibliography}
\end{document}